\documentclass[twoside,11pt]{article}

\usepackage{xcolor}
\usepackage[abbrvbib, preprint]{jmlr2e}
\usepackage{amsmath}
\usepackage{hyperref}
\usepackage[ruled, vlined]{algorithm2e}
\SetKwInput{KwInput}{Input}              
\SetKwInput{KwOutput}{Output} 

\newtheorem{assumption}[theorem]{Assumption}

\usepackage{lastpage}
\jmlrheading{22}{2021}{1-\pageref{LastPage}}{12/20; Revised
2/21}{4/21}{20-1462}{Steven Siwei Ye and Oscar Hernan Madrid Padilla}

\ShortHeadings{Non-parametric Quantile Regression via the K-NN Fused Lasso}{Ye and Padilla}
\firstpageno{1}

\begin{document}

\title{Non-parametric Quantile Regression via the K-NN Fused Lasso}

\author{\name Steven Siwei Ye \email stevenysw@g.ucla.edu \\
       \addr Department of Statistics\\
       University of California, Los Angeles\\
       Los Angeles, CA 90095, USA
       \AND
       \name Oscar Hernan Madrid Padilla \email oscar.madrid@stat.ucla.edu \\
       \addr Department of Statistics\\
       University of California, Los Angeles\\
       Los Angeles, CA 90095, USA}

\editor{Pradeep Ravikumar}

\maketitle

\begin{abstract}
Quantile regression is a statistical method for estimating conditional quantiles of a response variable. In addition, for mean estimation, it is well known that quantile regression is more robust to outliers than $l_2$-based methods. By using the fused lasso penalty over a $K$-nearest neighbors graph, we propose an adaptive quantile estimator in a non-parametric setup. We show that the estimator attains optimal rate of $n^{-1/d}$ up to a logarithmic factor, under mild assumptions on the data generation mechanism of the $d$-dimensional data. We develop algorithms to compute the estimator and discuss methodology for model selection. Numerical experiments on simulated and real data demonstrate clear advantages of the proposed estimator over state of the art methods. All codes that implement the algorithms and the datasets used in the experiments are publicly available on the author's Github page (\url{https://github.com/stevenysw/qt_knnfl}).
\\ \\
\end{abstract}

\section{Introduction}

\label{sec1}
Assume that we have $n$ observations, $(x_1,y_1),..., (x_n,y_n),$ of the pair of random variables $(X,Y)$. The response variable $Y$ is a real-valued vector and $X$ is a multivariate covariate or predictor variable in a metric space $\mathcal{X}$ with metric $d_{\mathcal{X}}$. A standard goal of non-parametric regression is to infer, in some way, the underlying relationship between $Y$ and $X$. The generative model behind this can be expressed as
\begin{equation}
\label{eq1}
    y_i = f_0(x_i) + \epsilon_i, \ \text{for } i = 1,...,n,
\end{equation}
where $f_0$ is an unknown function that we want to estimate. While usual regression considers estimation of the conditional mean of the response variable, quantile regression estimates the conditional median (or other desired quantiles) of the response variable. Specifically, given a quantile level $\tau \in (0,1)$, we can rewrite (\ref{eq1}) as 
\begin{equation}
\label{eq2}
y_i = \theta^{\ast}_i + \epsilon_i, \ \text{for } i = 1,...,n,
\end{equation}
where 
\begin{equation}
\label{eq3}
\theta^{\ast}_i = F^{-1}_{y_i|x_i}(\tau).
\end{equation}
Here, $\theta^{\ast}$ is the vector of $\tau$-quantiles of $y$, $F_{y_i|x_i}$ represents the cumulative distribution function of $y_i$ given $x_i$, and $\mathbb{P}(\epsilon_i \leq 0   |  x_i) = \tau$.

The goal of quantile regression is to estimate $\theta^{\ast}$ as accurately as possible, and it usually involves an optimization problem in the form 
\begin{equation}
\label{eq4}
\hat \theta \in \underset{\theta \in \zeta \subset \mathbb{R}^n}{\arg\min} \ L(\theta),
\end{equation}
where $L(\theta)$ is the loss function is defined by
\begin{equation}
\label{eq5}L(\theta) = \sum_{i=1}^n\rho_\tau(y_i-\theta_i),
\end{equation}
with $\rho_\tau(t) = (\tau -1\{t \leq 0\})t$, the asymmetric absolute deviation function \citep{Koenker78}.

In this paper, we apply total variation denoising for non-parametric quantile regression in a multivariate setup and combine it with the $K$-NN procedure. We leverage insights gained from recent results for quantile trend filtering \citep{Madrid20} and $K$-NN fused lasso \citep{Padilla20a}. Our proposed estimator, quantile $K$-NN fused lasso, can adapt to the piecewise linear or piecewise polynomial structure in the true vector properly.

It takes simply two steps to compute our proposed quantile $K$-NN fused lasso estimator. We first construct a $K$-nearest-neighbor graph corresponding to the given observations. The second step involves a constrained optimization problem with a lasso-type penalty along the $K$-NN graph as follow
\begin{equation}
\label{eqn:estimator}
\hat \theta \in \underset{\theta \in \zeta \subset \mathbb{R}^n}{\arg\min} \ L(\theta) + \lambda\Vert \nabla_G\theta\Vert_1,
\end{equation}
where $\lambda > 0$ is a tuning parameter. The notation $\nabla_G$ in the penalty term represents the oriented incidence matrix of the $K$-NN graph, and we will provide the detailed definition in Section \ref{sec2}.

We study the rate of convergence of the estimator defined in (\ref{eqn:estimator}). Towards that end, we denote a loss function $\Delta_n^2:\mathbb{R}^n \rightarrow \mathbb{R}$ by
$$\Delta_n^2(\delta):=\frac{1}{n}\sum_{i=1}^n \min\{|\delta_i|, \delta_i^2\}.$$
The function $\Delta_n^2(\cdot)$ appeared in \cite{Madrid20} and is similar to a Huber loss function \citep[see page 471 in][]{Wainwright19}, which offers a compromise between the least-squares cost and the $l_1$-norm cost function and is thus less sensitive to outliers in data. We show that under mild conditions, (\ref{eqn:estimator}) attains a convergence rate of $n^{-1/d}$ for $d$-dimensional data in terms of $\Delta_n^2$, ignoring the logarithmic factor. The rate is nearly minimax and it matches the mean squared error rates from \citet{Padilla20a}. However, unlike \citet{Padilla20a}, our result holds under general errors allowing for heavy-tailed distributions. Another notable point in our theoretical analysis, different from previous quantile regression work, is that we only require a bounded variation class of signals and hence guarantee our result to hold under very general models.

\subsection{Previous Work}
Quantile regression, since introduced by \citet{Koenker78}, has become a commonly used class of methods in many applications thanks to its flexibility and robustness for modelling conditional distributions. The study of quantile regression in non-parametric setups can be dated back to \citet{Utreras81}, \citet{Cox83} and \citet{Eubank88}, whose works were mainly developed for median regression on one-dimensional data. Later, \citet{Koenker94} proposed a more general estimator for any desired quantile $\tau$, quantile smoothing spline, and \citet{He98} provided a bivariate version of the estimator. Quantile smoothing spline problems are of a $l_1$ penalty structure, which led to a more general study on $l_1$-norm regularized quantile regression by \citet{Li08}. Other methods for non-parametric quantile regression have also been proposed in the literature. \citet{Yu98}, \citet{Cai98}, and \citet{Spokoiny13} explored local polynomial quantile regression. \citet{Belloni19} studied non-parametric series quantile regression, and \citet{Meinshausen06} introduced quantile random forests. Quantile regression with rectified linear unit (ReLU) neural networks is another edge-cutting approach that utilizes some knowledge from the field of deep learning, and the theory is studied in \citet{Padilla20b}.

Our approach exploits the local adaptivity of total variation denoising. \citet{Rudin92} first proposed total variation denoising for the application of image processing, and \citet{Tibshirani05} studied fused lasso thoroughly. Later, \citet{Kim09} extended the discussion to the so-called trend filtering on one-dimensional data, and \citet{Wang16} provided a generalization of trend filtering to the setting of estimation on graphs. These problems can be formulated similarly into an optimization problem of the form
\begin{equation}
\label{eq7}
\hat \theta \in \underset{\theta \in  \mathbb{R}^n}{\arg\min} \ \frac{1}{2}\sum_{i=1}^n (\theta_i-y_i)^2 + \lambda\Vert D\theta\Vert_1,
\end{equation}
where $\lambda >0$ is a regularization parameter to be chosen carefully and $D$ is a matrix. For instance, trend filtering order $k$ (fused lasso $k=0$) consists of matrices $D$ that capture the $(k+1)$th order total variation  of a given signal, see \citet{Tibshirani14}.  A substantial amount of literature focused on theoretical guarantees of these estimators. \citet{Mammen97} and \citet{Tibshirani14} showed that trend filtering  attains nearly minimax rates in mean squared error (MSE) for estimating functions of bounded variation. \citet{Huetter16} showed a sharp oracle rate of total variation denoising along grid graphs. More recently, \citet{Padilla20a} incorporated fused lasso with the $K$-NN procedure and proved that the $K$-NN fused lasso also achieves nearly minimax convergence rate. 
In the quantile regression literature, \citet{Belloni11}, \citet{Kato11}, \citet{Fan14} studied quantile model selection via $l_1$ regularization, but most of these works required strict linear assumptions. \citet{Madrid20} provided proofs for theoretical properties of quantile trend filtering estimator in one dimension. They showed that under minimal assumptions of the data generation mechanism, quantile trend filtering estimator attains a minimax rate for one-dimensional piecewise polynomial regression.

On the computational side, quantile regression is different from $l_2$ based methods because it requires a non-trivial reformulation of the optimization problem due to non-differentiability of the loss as (\ref{eq5}). The most well-known algorithm for computing a quantile estimator  is due to \citet{Koenker05} and  it uses an interior point (IP) approach. \citet{Pietrosanu17} studied high-dimensional quantile regression problems and obtained estimators by applying the alternating direction method of multipliers \citep[ADMM;][]{Boyd11}, majorize-minimize \citep[MM;][]{Hunter00}, and coordinate descent \citep[CD;][]{Wu08} algorithms for variable selection. For computing trend filtering estimates, \citet{Hochbaum17} developed a fast algorithm for quantile fused lasso in $O(n\log n)$ operations. Recently, \citet{Brantley20} proposed an ADMM based algorithm for computing $k$th order quantile trend filtering estimators for one-dimensional data.

\subsection{Outline of the Paper}
In Section \ref{sec2}, we provide the definition of quantile $K$-nearest-neighbors fused lasso estimator and the constrained version of the problem. Two algorithms to compute the proposed estimators numerically -- alternating directions method of multipliers (ADMM), and majorize-minimize (MM), are introduced in Section \ref{sec3}, and the discussion on how to select an appropriate penalty parameter in practice is also included. Section \ref{sec4} presents the two theoretical developments regarding the constrained and penalized estimators. The theorems demonstrate that under general assumptions, both estimators converge at a rate of $n^{-1/d}$, up to a logarithmic factor, for estimating $d$-dimensional data under the loss function $\Delta^2_n$ defined above. Section \ref{sec5} lists the results of numerical experiments on multiple simulated datasets and two real datasets, California housing data and Chicago crime data. The experiments show that the proposed estimator outperform state-of-the-art methods on both simulated and real datasets. Moreover, the comparison on accuracy and computational time among the algorithms introduced in Section \ref{sec3} provides the audience with some insights on choosing a suitable algorithm for specific problems. The proofs of theorems in the paper are provided in the Appendix. 

\section{Quantile K-NN Fused Lasso}
\label{sec2}
The first step to construct the quantile $K$-NN fused lasso estimator is to build a $K$-NN graph $G$. Specifically,  given the observations, $G$ has vertex set $V = \{1,...,n\}$, and its edge set $E_K$ contains the pair $(i,j)$, for $i \in V$, $j \in V$, and $i \neq j$, if and only if $x_i$ is among the $K$-nearest neighbors of $x_j$, with respect to the metric $d_{\mathcal{X}}$, or vice versa.  
After constructing the $K$-NN graph, we can formalize an optimization problem for quantile $K$-NN fused lasso as
\begin{equation}
\label{eq8}
 \hat \theta = \underset{\theta \in \mathbb{R}^n}{\arg\min} \ \sum_{i=1}^n \rho_\tau(y_i-\theta_i) + \lambda \Vert \nabla_{G} \theta\Vert_1,
\end{equation}
where $\lambda > 0$ is a tuning parameter, and $\nabla_G$ is an oriented incidence matrix of the $K$-NN graph $G$. Thus, we define $\nabla_G$ as follows: each row of the matrix corresponds to one edge in $G$; for instance, if the $p$-th edge in $G$ connects the $i$-th and $j$-th observations, then
$$\left(\nabla_G\right)_{p,q} = \begin{cases}
1 & \text{if } q = i,\\
-1 & \text{if } q = j,\\
0 & \text{otherwise.}
\end{cases}$$
In this way, the $p$-th element in $\nabla_G \theta$, $(\nabla_G \theta)_p = \theta_i - \theta_j$. Notice that we choose the ordering of the nodes and edges in $\nabla_G$ arbitrarily without loss of generality.

Once $\hat\theta$ in (\ref{eq8}) has been computed, we can predict the value of response corresponding to a new observation $x \in \mathcal{X}\backslash \{x_1, ...,x_n\}$ by the averaged estimated response of the $K$-nearest neighbors of $x$ in $\{x_1, ...,x_n\}$. Mathematically, we write
\begin{equation}
\label{eq9}
   \hat y = \frac{1}{K}\sum_{i=1}^n \hat\theta_i \cdot\textbf{1}\{x_i \in \mathcal{N}_K(x)\},
\end{equation}
where $\mathcal{N}_K(x)$ is the set of $K$-nearest neighbors of $x$ in the training data. A similar prediction rule was used in \citet{Padilla20a}.

A related estimator to the penalized estimator $\hat \theta$ defined in (\ref{eq8}) is the the constrained estimator $\hat \theta_C$, of which the corresponding optimization problem can be written as
\begin{equation} 
\label{eq10}
\begin{split}
 \hat \theta_C \ \  = \ \ &\underset{\theta \in \mathbb{R}^n}{\arg\min} \ \sum_{i=1}^n \rho_\tau(y_i-\theta_i) \\
    &\text{subject to} \ \ \Vert \nabla_{G} \theta\Vert_1 \leq C,
\end{split}
\end{equation}
for some positive constant $C$.

\subsection{Comparison with the K-NN fused lasso estimator }

Before proceeding to study the properties of our proposed method we provide some comparisons with its precursor, the $K$-NN fused lasso estimator from \citet{Padilla20a}. The latter is defined as

\begin{equation}
	\label{eqn:knnfl}
	\tilde{\theta}	\,= \,\underset{\theta \in \mathbb{R}^n}{\arg\min} \ \sum_{i=1}^n (y_i-\theta_i)^2 + \lambda \Vert \nabla_{G} \theta\Vert_1
\end{equation}
where  $\lambda>0$ is a tuning parameter.

In contrasting (\ref{eqn:knnfl}) with (\ref{eq8}) we first highly that from a practical perspective the latter has some important advantages. Firstly,  (\ref{eq8})   can be used  to construct  prediction intervals whereas (\ref{eqn:knnfl}) can only provide point estimates for the different locations. We illustrate the construction of prediction intervals with a real data example in Section \ref{sec5.2}. Secondly, the quantile $K$-NN fused lasso is by construction expected to be more robust to heavy tails and outliers than its counterpart the  $K$-NN fused lasso. We verify this in our experiments section.

On the computational side, the algorithms for solving the optimization problem in (\ref{eqn:knnfl})  cannot be used  for finding the estimator in (\ref{eq8}). Hence, novel algorithms are needed to efficiently compute our proposed estimator. This is the subject of the next section.

Finally,  despite the similarity in the definition of  the estimators in  (\ref{eqn:knnfl}) and (\ref{eq8}), the theory from \cite{Padilla20a} does not directly translate to analyze our estimator defined in (\ref{eq8}). Hence, one of the contributions of this paper is to show that the quantile $K$-NN fused lasso   inherits local adaptivity properties of the  $K$-NN fused lasso in general settings that allow for heavy-tailed error distributions.

\section{Algorithms and Model Selection}
\label{sec3}
To compute the quantile $K$-NN fused lasso estimator, the first step is to construct the $K$-NN graph from the data. The computational complexity of constructing the $K$-NN graph is of $O(n^2)$, although it is possible to accelerate the procedure to $O(n^t)$ for some $t\in (1,2)$ using divide and conquer methods \citep[][]{Bentley80, Chen09}. 

The second step of computation is to solve a constrained optimization problem as (\ref{eq8}). Here, we introduce three algorithms to solve the problem numerically. Before presenting our algorithms, we stress that both the problems (\ref{eq8}) and (\ref{eq10}) are linear programs, therefore we can use any linear programming software to obtain an optimal solution. Noticeably, we can take the advantage of sparsity in the penalty matrix  for faster computation. However, a shortcoming of linear programming is that the algorithm can become very time-consuming for large sized problems, especially when $n$ is greater than 5000.

\subsection{Alternating Directions Method of Multipliers (ADMM)}
\label{sec3.1}
The alternating directions method of multipliers (ADMM) algorithm \citep{Boyd11} is a powerful tool for solving constrained optimization problems.

We first reformulate the optimization problem (\ref{eq8}) as
\begin{equation}
\label{eq11}
\begin{split}
    &\underset{\theta\in \mathbb{R}^n, z \in \mathbb{R}^n}{\text{minimize}} \ \sum_{i=1}^n \rho_\tau(y_i-\theta_i) + \lambda \Vert \nabla_{G} z\Vert_1 \\
    &\text{s.t. } \ \ \ \ \ \ \ \ \ \ \ z= \theta,
\end{split}
\end{equation}
and the augmented Lagrangian can then be written as
$$L_R(\theta,z,u) =  \sum_{i=1}^n \rho_\tau(y_i-\theta_i) + \lambda \Vert \nabla_{G} z\Vert_1 + \frac{R}{2}\Vert \theta - z + u\Vert^2,$$
where $R$ is the penalty parameter that controls step size in the update. Thus we can solve (\ref{eq11}) by iteratively updating the primal and dual
\begin{equation}
\label{eq12}
    \ \ \ \ \theta \leftarrow \underset{\theta\in \mathbb{R}^n}{\arg\min} \ \left\{\sum_{i=1}^n\rho_\tau(y_i-\theta_i) + \frac{R}{2}\Vert\theta - z + u\Vert^2\right\},
\end{equation}
\begin{equation}
\label{eq13}
    z \leftarrow \underset{z\in \mathbb{R}^n}{\arg\min} \ \left\{ \frac{1}{2}\Vert \theta + u -z\Vert^2 + \frac{\lambda}{R}\Vert \nabla_G z\Vert_1 \right\}.
\end{equation}
The primal problem (\ref{eq12}) can be solved coordinate-wise in closed form as
\begin{equation*}
\theta_i = \begin{cases}
z_i - u_i + \frac{\tau}{R} & \text{ if } y_i - z_i + u_i > \frac{\tau}{R},\\
z_i - u_i + \frac{\tau-1}{R} & \text{ if } y_i - z_i+ u_i < \frac{\tau-1}{R},\\
y_i & \text{ otherwise; }
\end{cases}
\end{equation*}
See Appendix A for the steps to derive the solution. The dual problem (\ref{eq13}) is a generalized lasso problem that can be solved with the parametric max-flow algorithm from \citet{Chambolle09}. 

The entire procedure is presented in Algorithm \ref{alg1}. In practice, we can simply choose the penalty parameter $R$ to be $\frac{1}{2}$. We require the procedure to stop if it reaches the maximum iteration or the primal residual $\Vert \theta^{(k)}-\theta^{(k-1)}\Vert_2$ is within  a tolerance $\kappa$, to which we set $10^{-2}$ in our computation. Actually, the ADMM algorithm converges very quickly with only tens of iterations and hence we find it to be faster than linear programming.  Another advantage of ADMM is that the algorithm is not sensitive to the initialization. In Appendix B, we present a simulation study to demonstrate the fast convergence of ADMM under different initializations.

\begin{algorithm}[h!]
  \KwInput{Number of nearest neighbor: $K$, quantile: $\tau$, penalty parameter: $\lambda$, maximum iteration: $N_{\text{iter}}$, tolerance: $\kappa$}
  \KwData{$X \in \mathbb{R}^{n\times d}$, $y \in \mathbb{R}^n$}
  \KwOutput{$\hat\theta \in \mathbb{R}^n$}
\textbf{1.} Compute $K$-NN graph incidence matrix $\nabla_G$ from $X$.\\
\textbf{2.} Initialize $\theta^{(0)} = y, z^{(0)} = y, u^{(0)} = 0$.\\
\textbf{3.} For $k= 1, 2, ...,$ until $\Vert \theta^{(k)}-\theta^{(k-1)}\Vert_2 \leq \kappa$ or the procedure reaches $N_\text{iter}$:\\
\ \ (a) For $i=1,...,n$, update $$\theta^{(k)}_i \leftarrow {\arg\min} \left\{ \rho_\tau(y_i-\theta_i) + \frac{R}{2}(\theta_i - z^{(k-1)}_i + u^{(k-1)}_i)^2\right\}.$$
(b) Update $$z^{(k)} \leftarrow {\arg\min} \left\{\frac{1}{2}\Vert \theta^{(k)} + u^{(k-1)} - z\Vert^2 + \frac{\lambda}{R}\Vert \nabla_G z\Vert_1\right\}.$$
(c) Update \begin{center}$u^{(k)} \leftarrow  u^{(k-1)} + \theta^{(k)} - z^{(k)}.$\end{center}
\caption{Alternating Directions Method of Multipliers for quantile $K$-NN fused lasso}
\label{alg1}
\end{algorithm}

\subsection{Majorize-Minimize (MM)}
\label{sec3.2}
We now exploit the majorize-minimize (MM) approach from \citet{Hunter00} for estimating the conditional median. The main advantage of the MM algorithm versus ADMM is that it is conceptually simple and it is a descent algorithm.  

Recall that for $\tau = 0.5$, quantile regression becomes least absolute deviation, or $L_1$-regression, and the loss function $L(\theta)$ of our problem becomes
\begin{equation}
\label{eq14}
 L(\theta) = \sum_{i=1}^n |y_i-\theta_i|+ \lambda \Vert \nabla_{G} \theta\Vert_1.
\end{equation}
Next, notice that $L(\theta)$ can be majorized at $\theta^{(k)}$ by $Q(\theta \ | \ \theta^{(k)})$ given as
\begin{equation}
\label{eq15}
 Q(\theta \ | \ \theta^{(k)}) =  \sum_{i=1}^n \frac{(y_i-\theta_i)^2}{|y_i-\theta^{(k)}_i|}+ \lambda \sum_{(i,j) \in E_K}\frac{(\theta_i-\theta_j)^2}{|\theta^{(k)}_i-\theta^{(k)}_j|} + \text{const.},
\end{equation}
since it holds that 
\begin{equation}
\label{eq16}
\begin{split}
    &Q(\theta^{(k)}\ | \ \theta^{(k)}) \ \ = \ \ L(\theta^{(k)}),\\
    &Q(\theta\ | \ \theta^{(k)}) \ \ \ \ \ \ \geq \ \ L(\theta) \ \text{ for all } \theta. 
\end{split}
\end{equation}
To avoid possible occurrences of zero, we add a perturbation $\epsilon$ to the denominator each time. Then, the iterative algorithm optimizes $Q(\theta \ | \ \theta^{(k)})$ at each iteration. The stopping criterion for MM algorithm remains the same as for ADMM. Because the optimization problem here has a closed-form solution, we can compute the solution directly by solving a linear system (see Step 3c in Algorithm \ref{alg2}). We find the MM algorithm to be faster in running time than linear programming and ADMM for large-size problems and it produces reasonably stable solutions as the others; see the experiments and discussion in Section \ref{sec5.1}. A major drawback of our fast algorithm is that it can only handle median regression at this moment. We leave for future work studying an extension of an MM-based algorithm for estimating general quantiles in the future.

\begin{algorithm}[h!]
  \KwInput{Number of nearest neighbor: $K$,  penalty parameter: $\lambda$, maximum iteration: $N_{\text{iter}}$,  tolerance: $\kappa$ }
  \KwData{$X \in \mathbb{R}^{n\times d}$, $y \in \mathbb{R}^n$}
  \KwOutput{$\hat\theta \in \mathbb{R}^n$}
\textbf{1.} Compute $K$-NN graph incidence matrix $\nabla_G$ from $X$.\\
\textbf{2.} Initialize $\theta^{(0)}_i = \text{median}(y)$ for $i=1,...,n$.\\
\textbf{3.} For $k= 1, 2, ...,$ until $\Vert \theta^{(k)}-\theta^{(k-1)}\Vert_2 \leq \kappa$ or the procedure reaches $N_\text{iter}$:\\
\ \ (a) Compute weight matrix $W \in \mathbb{R}^{n\times n}$:
$$W = \text{diag}(1/[|y-\theta^{(k-1)}|+\epsilon]).$$
(b) Compute weight matrix $$\tilde W = \text{diag}(1/[|\theta^{(k-1)}_i-\theta^{(k-1)}_j|+\epsilon]) \ \  \text{if }(i,j) \in E_K.$$
(c) Update 
\begin{center}$\theta^{(k)} \leftarrow [W + \lambda \nabla_G^\top \tilde W\nabla_G]^{-1}Wy$.
\end{center}
\caption{Majorize-Minimize for quantile $K$-NN fused lasso, $\tau = 0.5$}
\label{alg2}
\end{algorithm}

\subsection{Model Selection}
\label{sec3.3}
The choice of the tuning parameter $\lambda$ in (\ref{eq8}) is an important practical issue in estimation because it controls the degree of smoothness in the estimator. The value of $\lambda$ can be chosen through K-fold cross-validation. Alternatively, we can select the regularization parameter based on Bayesian Information Criteria \citep[BIC;][]{Schwarz78}. The BIC for quantile regression \citep{Keming01} can be computed as 
$$\text{BIC}(\tau) = \frac{2}{\sigma}\sum_{i=1}^n\rho_\tau(y_i-\hat\theta_i) + \nu\log n,$$
where $\nu$ denotes the degree of freedom of the estimator and $\sigma >0$ can be empirically chosen as $\sigma = \frac{1-|1-2\tau|}{2}$. It is also possible to use Schwarz Information Criteria \citep[SIC;][]{Koenker94} given by
$$\text{SIC}(\tau) = \log\left[\frac{1}{n}\sum_{i=1}^n\rho_\tau(y_i-\hat\theta_i)\right] + \frac{1}{2n}\nu\log n.$$
However, BIC is more stable than SIC in practice because when $\lambda$ is small, SIC may become ill-conditioned  as the term inside the logarithm is close to 0.

\citet{Tibshirani12} demonstrate that a lasso-type problem has degrees of freedom
$\nu$ equal to the expected nullity of the penalty matrix after removing the rows that are indexed by the boundary set of a dual solution at $y$. In our case, we define $\nu$ as the number of connected components in the graph $\nabla_G$ after removing all edges $j$'s where $|(\nabla_G\hat\theta )_j|$ is above a threshold $\gamma$, typically very small (e.g., $10^{-2}$).

\section{Theoretical Analysis}
\label{sec4}
Before arriving at our main results, we introduce some notation. For a set $A \subset \mathcal{A}$ with $(\mathcal{A}, d_{\mathcal{A}})$ a metric space, we write $B_\epsilon(A) = \{a:\text{exists } a' \in A, \text{ with } d_{\mathcal{A}}\leq \epsilon\}$. The Euclidean norm of a vector $x \in \mathbb{R^d}$ is denoted by $\Vert x\Vert_2 = (x_1^2+...+x^2_d)^{1/2}$. The $l_1$ norm of $x$ is denoted by $\Vert x\Vert_1 = |x_1| + ... + |x_d|$. The infinity norm of $x$ is denoted by $\Vert x\Vert_\infty = \max_i |x_i|$. In the covariate space $\mathcal{X}$, we consider the Borel sigma algebra, $\mathcal{B}(\mathcal{X})$, induced by the metric $d_{\mathcal{X}}$, and we let $\mu$ be a measure on $\mathcal{B}(\mathcal{X})$. We assume that the covariates in the model (\ref{eq1}) satisfy $x_i\overset{\text{ind}}{\sim} p(x)$. In other word, $p$ is the probability density function associated with the distribution of $x_i$, with respect to the measure space $(\mathcal{X}, \mathcal{B}(\mathcal{X}), \mu)$. Let $\{a_n\}$ and $\{b_n\} \subset \mathbb{R}$ be two sequences. We write $a_n = O_{\mathbb{P}}(b_n)$ if for every $\epsilon > 0$ there exists $C > 0$ such that $\mathbb{P}(a_n \geq Cb_n) < \epsilon$ for all $n$. We also write $\textup{poly}(x)$ as a polynomial function of $x$, but the functions vary from case to case in the content below. Throughout, we assume the dimension of $\mathcal{X}$ to be greater than 1, since the study of the one-dimensional model, quantile trend filtering, can be found in \citet{Madrid20}.

We first make some necessary assumptions for analyzing the theoretical guarantees of both the constrained and penalized estimators.

\begin{assumption}
\label{as1}
(Bounded Variation) We write $ \theta^*_i = F^{-1}_{y_i|x_i}(\tau)$ for $i = 1,...,n$ and require that $V^*:= \Vert \nabla_{G} \theta^*\Vert_1 /[n^{1-1/d}\textup{poly}(\log n)]$ satisfies $V^*=O_\mathbb{P}(1)$. Here $F_{y_i|x_i}$ is the cumulative distribution function of $y_i$ given $x_i$ for $i = 1,...,n$.
\end{assumption}

The first assumption simply requires that $\theta^*$, the vector of $\tau$-quantiles of $y$, has bounded total variation along the $K$-NN graph. The scaling $n^{1-1/d}\textup{poly}(\log n)$ comes from \citet{Padilla20a}, and we will discuss the details after we present Assumptions \ref{as3}--\ref{as5} below.

\begin{assumption}
\label{as2} 
 There exists a constant $L>0$ such that for $\delta \in \mathbb{R}^n$ satisfying $\Vert \delta \Vert_\infty \leq L$ we have that
$$\min_{i=1,...,n} f_{y_i|x_i}(\theta^*_i + \delta_i) \geq \underline{f} \ \ \ \text{  a.s.},$$
for some $\underline{f} > 0$, and where $f_{y_i|x_i}$ is the the conditional probability density of $y_i$ given $x_i$.
\end{assumption}

The assumption that the conditional density of the response variable is bounded by below in a neighborhood is standard in quantile regression analysis. Related conditions appears as D.1 in \citet{Belloni11} and Condition 2 in \citet{Xuming94}.

The next three assumptions inherit from the study of $K$-NN graph in \citet{vonLuxburg14} and \citet{Padilla20a}. 

\begin{assumption}
\label{as3}
The density of the covariates, $p$, satisfies $0 < p_{\min} < p(x) < p_{\max}$, for all $x \in \mathcal{X}$, where $p_{\min}, p_{\max} > 0$ are fixed constants.
\end{assumption}

We only require the distribution of covariates to be bounded above and below by positive constants. In \citet{Gyorfi06} and \citet{Meinshausen06}, $p$ is assumed to be the probability density function of the uniform distribution in $[0,1]^d$.

\begin{assumption}
\label{as4}
The base measure $\mu$ in the metric space $\mathcal{X}$, in which $X$ is defined, satisfies $$c_{1,d}r^d \leq \mu\{B_r(x)\}\leq c_{2,d}r^d, \ \text{for all } x \in \mathcal{X},$$
for all $0 < r<r_0,$ where $r_0, c_{1,d}, c_{2,d}$ are positive constants, and $d \in \mathbb{N} \backslash \{0,1\}$ is the intrinsic dimension of $\mathcal{X}$.
\end{assumption}

Although $\mathcal{X}$ is not necessarily a Euclidean space, we require in this condition that balls in $\mathcal{X}$ have volume, with respect to some measure $\mu$ on the
Borel sigma algebra, $\mathcal{B}(\mathcal{X})$, that behaves similarly to the Lebesgue measure of balls in $\mathbb{R}^d$.

\begin{assumption}
\label{as5}
There exists a homeomorphism $h:\mathcal{X} \rightarrow [0,1]^d$, such that
$$L_{\min} d_\mathcal{X}(x,x') \leq \Vert h(x)-h(x')\Vert_2 \leq L_{\max} d_\mathcal{X}(x,x'),$$
for all $x,x' \in \mathcal{X}$ and for some positive constants $L_{\min},L_{\max}$, where $d \in \mathbb{N} \backslash \{0,1\}$ is the intrinsic dimension of $\mathcal{X}$.
\end{assumption}

The existence of a continuous bijection between $\mathcal{X}$ and $[0,1]^d$ ensures that the space has no holes and is topologically equivalent to $[0,1]^d$. Furthermore, Assumption \ref{as5} requires   $d>1$. If  $d =1$ then one can simply order the covariates and then run the one dimensional quantile fused lasso studied in \cite{Madrid20}. 

On another note, we point that \citet{Padilla20a} showed, exploiting ideas from \cite{vonLuxburg14},  that under Assumptions \ref{as3}--\ref{as5},
$\Vert \nabla_G\theta^* \Vert_1 \asymp n^{1-1/d}$ for a $K$-NN graph $G$ up to a polynomial of $\log n$, with an extra condition on the function $f_0 \circ h^{-1}$ being piecewise Lipschitz. Hence, under such conditions Assumption \ref{as1} holds. For completeness, the definition of the class of piecewise Lipschitz functions is provided in Appendix E. It is rather remarkable that \citet{Padilla20a} also present an alternative condition on $f_0 \circ h^{-1}$ than piecewise Lipschitz to guarantee the results hold; see Assumption 5 in the same paper.

Now, we are ready to present our first theoretical result on quantile $K$-NN fused lasso estimates.
\begin{theorem}
\label{theorem1}
Under Assumptions \ref{as1}--\ref{as5}, by setting $C = \frac{V}{n^{1-1/d}}$ in (\ref{eq10}) for a tuning parameter $V$, where $V \asymp 1$ and $V \geq V^{\ast}$, we have
$$\Delta_n^2\left(\theta^{\ast}-\hat\theta_C\right)=O_{\mathbb{P}}\left\{n^{-1/d}\textup{poly}(\log n)\right\},$$
for a choice of  $K$  satisfying  $K = \textup{poly}(\log n)$.
\end{theorem}

The first theorem shows that quantile $K$-NN fused lasso attains the optimal rate of $n^{-1/d}$ under the loss $\Delta^2_n(\cdot)$ defined in Section \ref{sec1} for estimating signals in a constrained set. 

\begin{theorem}
\label{theorem2}
Under Assumptions \ref{as1}--\ref{as5}, there exists a choice of $\lambda$ for (\ref{eq8}) satisfying
$$\lambda = \begin{cases}\Theta\left\{\log n\right\} & \text{ for } d = 2,\\
\Theta\left\{(\log n)^{1/2}\right\} & \text{ for } d > 2,\\
\end{cases}$$
such that 
$$\Delta_n^2\left(\theta^{\ast}-\hat\theta\right)= O_{\mathbb{P}}\left\{n^{-1/d}\textup{poly}(\log n)\right\},
$$
for a choice of  $K$  satisfying  $K = \textup{poly}(\log n)$.
\end{theorem}
The second theorem states that, under certain choice of the tuning parameter, the penalized estimator achieves the convergence rate of $n^{-1/d}$, similar to the constrained estimator, ignoring the logarithmic factor.   Notice that there is a one-to-one correspondence between (\ref{eq8}) and  (\ref{eq10}), as they are equivalent optimization problems.  Let $\lambda>0$   as in the statement of  Theorem  \ref{theorem2}. Then there exists a $C$ that depends on $y$    and  $\lambda$ such that  (\ref{eq8}) and  (\ref{eq10}) have identical solutions. However, the fact that such $C$  depends on $y$ does not imply that  for  a deterministic $C$, such as that in Theorem \ref{theorem1},  one can conclude that  if the unconstrained estimator  attains the rate $n^{-1/d}$  then the  constrained estimator attains the same rate. Nevertheless, the fact that we have  Theorems \ref{theorem1}--\ref{theorem2} implies that both estimators attain the same rate.

We emphasize that if all or some of  Assumptions \ref{as1}, \ref{as3}--\ref{as5}   do not hold, then  we are not able to characterize the behavior of the $K$-NN graph. In such case the conclusions of Theorems \ref{theorem1}--\ref{theorem2} would not necessarily hold.

We conclude with a remark regarding the minimax optimality of Theorems \ref{theorem1}--\ref{theorem2}.

\begin{remark}
	Let   $f^*(t) =  F^{-1}_{Y|X =t }(0.5)$ be the median function  and  let $\mathcal{C}$ be the class of piecewise Lipschitz functions, see Definition  \ref{def_lip} in Appendix E. It was proven in Proposition 2  of  \cite{castro2005faster} that  under the assumption that     $\epsilon_i \overset{\text{ind}}{\sim} N(0,\sigma^2)$ for some fixed  $\sigma$, and that $x_i \overset{\text{ind}}{\sim} U([0,1]^d)$ for  $i=1,\ldots,n$, it holds that
	\begin{equation}
		\label{eqn:lower1}
		\underset{ \hat{f} \,\,\text{estimator}  }{\inf }\,\,\,\underset{f^*  \in \mathcal{C},\,\,\, \|f^* \|_{\infty} \leq 1       }{\sup}   \mathbb{E}\left[   \int_{[0,1]^d}     (f^*(t) - \hat{f}(t)) ^2dt      \right] \,\geq\, c n^{-1/d},
	\end{equation}
	for some constant $c>0$.  However, by the constraint  $\|f^* \|_{\infty} \leq 1   $, the left hand side of (\ref{eqn:lower1})   equals, up to a constant, to
	\begin{equation}
		\label{eqn:lower2}
		\underset{ \hat{f} \,\,\text{estimator}   }{\inf }\,\,\,\underset{f^*  \in \mathcal{C},\,\,\, \|f^* \|_{\infty} \leq 1       }{\sup}   \mathbb{E}\left[   \int_{[0,1]^d}    \min\{ \vert f^*(t) - \hat{f}(t)  \vert ,  (f^*(t) - \hat{f}(t)) ^2 \}dt      \right].
	\end{equation}
	Furthermore, a discrete version of   
	\[
	\mathbb{E}\left[   \int_{[0,1]^d}    \min\{ \vert f^*(t) - \hat{f}(t)  \vert ,  (f^*(t) - \hat{f}(t)) ^2 \}dt      \right]
	\]
	is the quantity $\Delta_n^2(\hat{\theta} -\theta^*)$    if  $\hat{\theta}_i = \hat{f}(x_i)$ and  $\theta_i^* = f^*(x_i)$ for $i=1,\ldots,n$. Therefore, the rates in Theorems \ref{theorem1}--\ref{theorem2} are nearly minimax in the sense that they match, up to log factors,  the lower bound  $n^{-1/d}$ on the quantity   (\ref{eqn:lower2}) without requiring sub-Gaussian errors, and under more general conditions on the covariates than uniform draws. In contrast, 
	under sub-Gaussian errors, the $K$-NN  fused lasso estimator  from  \cite{Padilla20a}  attains, up to log factors,  the rate  $n^{-1/d}$  in terms of  mean squared error.

\end{remark}

\section{Experiments}
\label{sec5}
In this section, we will examine the performance of quantile $K$-NN fused lasso (QKNN) on various simulated and real datasets. The two benchmark estimators we compare against are $K$-NN fused lasso \citep[KNN;][]{Padilla20a} and quantile random forest \citep[QRF;][]{Meinshausen06}. The performance of an estimator is measured by its mean squared error, defined by
$$\text{MSE}(\hat\theta):=\frac{1}{n}\sum_{i=1}^n (\hat\theta_i-\theta_i^*)^2,$$
where $\theta^*$ is the vector of $\tau$-quantiles of the true signal.

For quantile $K$-NN fused lasso, we use the ADMM algorithm and select the tuning parameter $\lambda$ based on the BIC criteria described in Section \ref{sec3.3}; for $K$-NN fused lasso, we use the algorithm from \citet{Chambolle09} and the corresponding penalty parameter is chosen to minimize the average mean squared error over 500 Monte Carlo replicates. For quantile random forest, we directly use the R package “quantregForest” with defaulted choice of tree structure and tuning parameters.

Throughout, for both $K$-NN fused lasso and quantile $K$-NN fused lasso, we set $K$ to be 5 for sufficient information and efficient computation.

\subsection{Simulation Study}
\label{sec5.1}
We generate 500 data sets from models under each scenario described below with sample size between $10^2$ and $10^4$ and then report the mean squared errors of the three estimators with respect to different quantiles. For each scenario the data are generated as
$$y_i = \theta^*_i + \epsilon_i, \text{ and } \theta^*_i = f_0(x_i), \ i = 1,...,n,$$
where $\theta^*_i$ comes from some underlying functions $f_0$, and the errors $\{\epsilon_i\}_{i=1}^n$ are independent with $\epsilon_i \sim F_i$ for some distributions $F_i$, where we select from Gaussian, Cauchy, and $t$-distributions. 

\newpage
\noindent\textbf{Scenario 1} \\
We generate $x_i$ uniformly from $[0,1]^2$, and define $f_0:[0,1]^2 \rightarrow \{0,1\}$ by
$$
f_0(x) = \begin{cases}
1 & \text{if } \ \frac{5}{4}x_{i1} +\frac{3}{4} x_{i2} > 1,\\ 
0 & \text{otherwise.}
\end{cases}
$$
\noindent\textbf{Scenario 2} \\
In this case, we generate $X \in \mathbb{R}^2$ according to the probability density function
\begin{equation*}
\begin{split}
    p(x) &= \frac{1}{5} \textbf{1}_{\{[0,1]^2\backslash[0.4,0.6]^2\}}(x) + \frac{16}{25} \textbf{1}_{\{[0.45,0.55]^2\}}(x) \\
    & \ \ \ +\frac{4}{25}\textbf{1}_{\{[0.4,0.6]^2 \backslash [0.45,0.55]^2\}}(x).
\end{split}
\end{equation*}
The function $f_0: \ [0,1]^2 \rightarrow \mathbb{R}$ is defined as
$$f_0(x) = \textbf{1}_{\{\Vert x - \frac{1}{2}(1,1)^\top \Vert^2_2 \leq \frac{2}{1000}\}}(x).$$
\textbf{Scenario 3} \\
Again, $x_i$ are from uniform $[0,1]^2$. The smooth function $f_0: \ [0,1]^2 \rightarrow \mathbb{R}$ is defined as
\begin{equation*}
    f_0(x_i) = 0.4x_{i1}^2 + 0.6x_{i2}^2.
\end{equation*}
\textbf{Scenario 4} \\
The function $f_0: \ [0,1]^d \rightarrow \mathbb{R}$ is defined as
$$f_0(x) = \begin{cases}
1 & \text{if } \ \Vert x - \frac{1}{4}1_d\Vert_2 < \Vert x - \frac{3}{4}1_d\Vert_2,\\ 
-1 & \text{otherwise,}
\end{cases}$$
and the density $p$ is uniform in $[0,1]^d$. The errors are chosen as $(x^\top \beta)\epsilon$, where $\beta = (\frac{1}{d},...,\frac{1}{d})^\top$. Here we simulate with $d= 5$. \\

The scenarios above have been chosen to illustrate the local adaptivity of our proposed approach to discontinuities of the quantile function. Scenario 1 consists of a piecewise constant median function. Scenario 2 is borrowed from \cite{Padilla20a} and also has a piecewise constant median. However, the covariates in Scenario 2 are not uniformly drawn and are actually highly  concentrated in a small region of the domain. Scenario 3 is a  smooth function, and Scenario 4 is taken from \cite{Padilla20a}. In the latter we also have a piecewise constant quantiles, but the boundaries of  the different pieces are not axis aligned, and the errors are heteroscedastic.

\begin{table*}[h!]
    \centering
        \begin{tabular}{c c c c c c c c}
        \hline
        n & Scenario & $\epsilon$ & $\tau$ & QKNN & QRF & KNN\\
        \hline
        100 & 1 & N(0,1) & 0.5 & 0.2364 (0.0628) & 0.2242 (0.0467) & \textbf{0.1507} (0.0365)\\
         1000 & 1 & N(0,1) & 0.5 & 0.1225 (0.0117) & 0.1569 (0.0116) & \textbf{0.0872} (0.0239)\\
         5000 & 1 & N(0,1) & 0.5 & 0.0983 (0.0024) & 0.1348 (0.0047) & \textbf{0.0540} (0.0022) \\
         10000 & 1 & N(0,1) & 0.5 & 0.0370 (0.0019) & 0.1279 (0.0028) & \textbf{0.0292} (0.0010) \\
         100 & 1 & Cauchy(0,1) & 0.5 & \textbf{0.1776} (0.0746) & 3059.63 (54248) & 35.4988 (124.19) \\
         1000 & 1 & Cauchy(0,1) & 0.5 &  \textbf{0.1440} (0.0817)& 26640.10 (179298) & 2816.1 (7691.6)\\
         5000 & 1 & Cauchy(0,1) & 0.5 & \textbf{0.1326} (0.0740) & 34038.86 (196566) & 4628.8 (7989.0)\\
         10000 & 1 & Cauchy(0,1) &  0.5& \textbf{0.0962} (0.0642) & 57748.95 (200327) & 4799.0 (7249.6)\\
         \hline
         100 & 2 & $t_3$ & 0.5 & \textbf{0.1838} (0.0467) & 0.5591 (0.3569) & 0.2374 (0.0595) \\ 
         1000 & 2 & $t_3$ & 0.5 & \textbf{0.0780} (0.0135) & 0.3935 (0.0875) & 0.1428 (0.0536)\\
         5000 & 2 & $t_3$ & 0.5 & \textbf{0.0364} (0.0050) & 0.3479 (0.0477) & 0.0622 (0.0165)\\
         10000 & 2 & $t_3$ & 0.5 & \textbf{0.0265} (0.0024) & 0.3311 (0.0457) & 0.0542 (0.0097)\\
         \hline
         100 & 3 & $t_2$ & 0.5 & \textbf{0.0470} (0.0253)& 1.6723 (4.8880) & 0.1545 (0.4188) \\
         1000 & 3 & $t_2$ & 0.5 & \textbf{0.0174} (0.0050) & 1.5363 (2.9381) & 0.0510 (0.0230)\\
         5000 & 3 & $t_2$ & 0.5 & \textbf{0.0075} (0.0015) & 1.4207 (2.1827) & 0.0414 (0.0106) \\
         10000 & 3 & $t_2$ & 0.5 & \textbf{0.0059} (0.0019) & 1.2910 (1.1924) & 0.0409 (0.0102) \\
         \hline
         100 & 4 & $t_3$ & 0.9 & 0.7413 (0.2180) & \textbf{0.6948} (0.5340) & * \\
         1000 & 4 & $t_3$ & 0.9 & \textbf{0.3568} (0.1192) & 0.5374 (0.2656) & * \\
         5000 & 4 & $t_3$ & 0.9 & \textbf{0.2889} (0.0683) & 0.4420 (0.0422) & * \\
         10000 & 4 & $t_3$ & 0.9 & \textbf{0.2409} (0.0173) & 0.4344 (0.0548) & * \\
         100 & 4 & $t_3$ &  0.1 & 1.1563 (0.3923) & \textbf{0.8877} (0.6858) & * \\
         1000 & 4 & $t_3$ & 0.1 & \textbf{0.4160} (0.1542) & 0.6224 (0.2667) & * \\
         5000 & 4 & $t_3$ & 0.1 & \textbf{0.3074} (0.0503) & 0.4897 (0.0644) & * \\
         10000 & 4 & $t_3$ & 0.1 & \textbf{0.2597} (0.0301)& 0.4536 (0.0494) & * \\
         \hline
    \end{tabular}
    \caption{Mean squared error $\frac{1}{n}\sum_{i=1}^n (\hat\theta_i-\theta_i^*)^2$, averaging over 500 Monte Carlo simulations for the different methods, sample sizes, errors, quantiles considered. The numbers in parentheses indicate the standard Monte Carlo errors over the replications.}
    \label{table:1}
\end{table*}

\begin{figure}[hbp!]
\centering
\includegraphics[width = 152mm, height = 44.2mm]{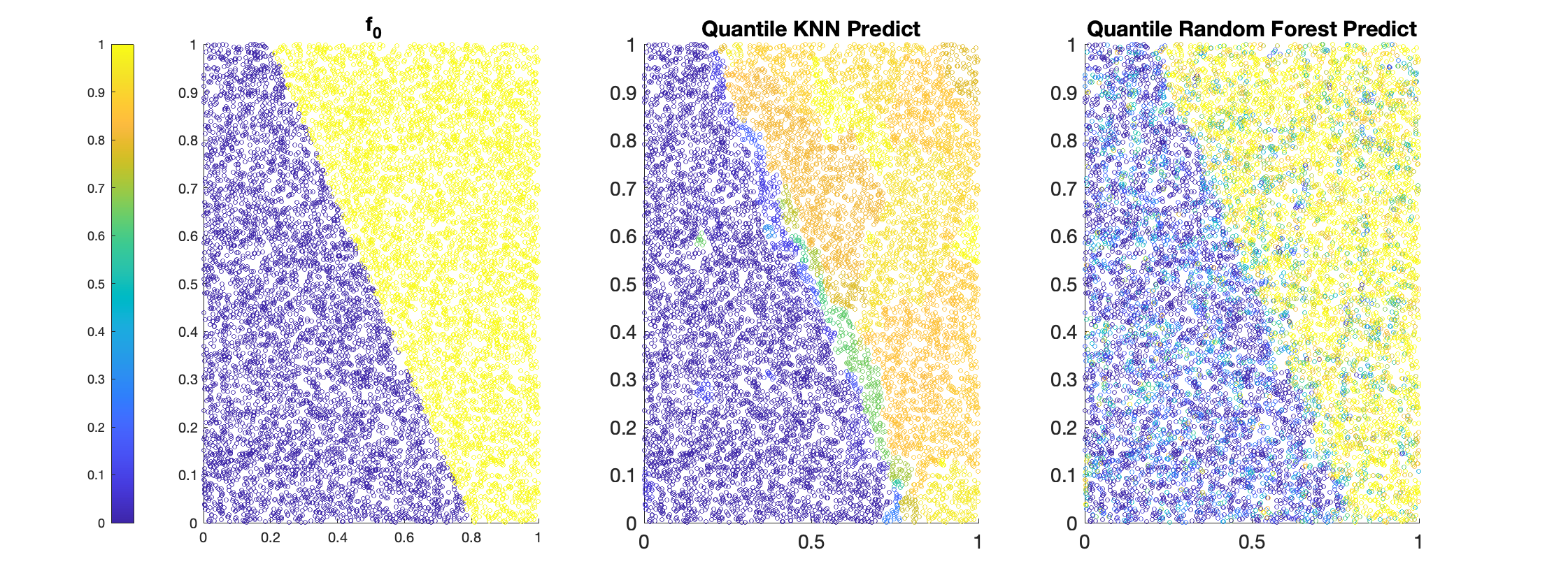}\\
\includegraphics[width = 152mm, height = 44.2mm]{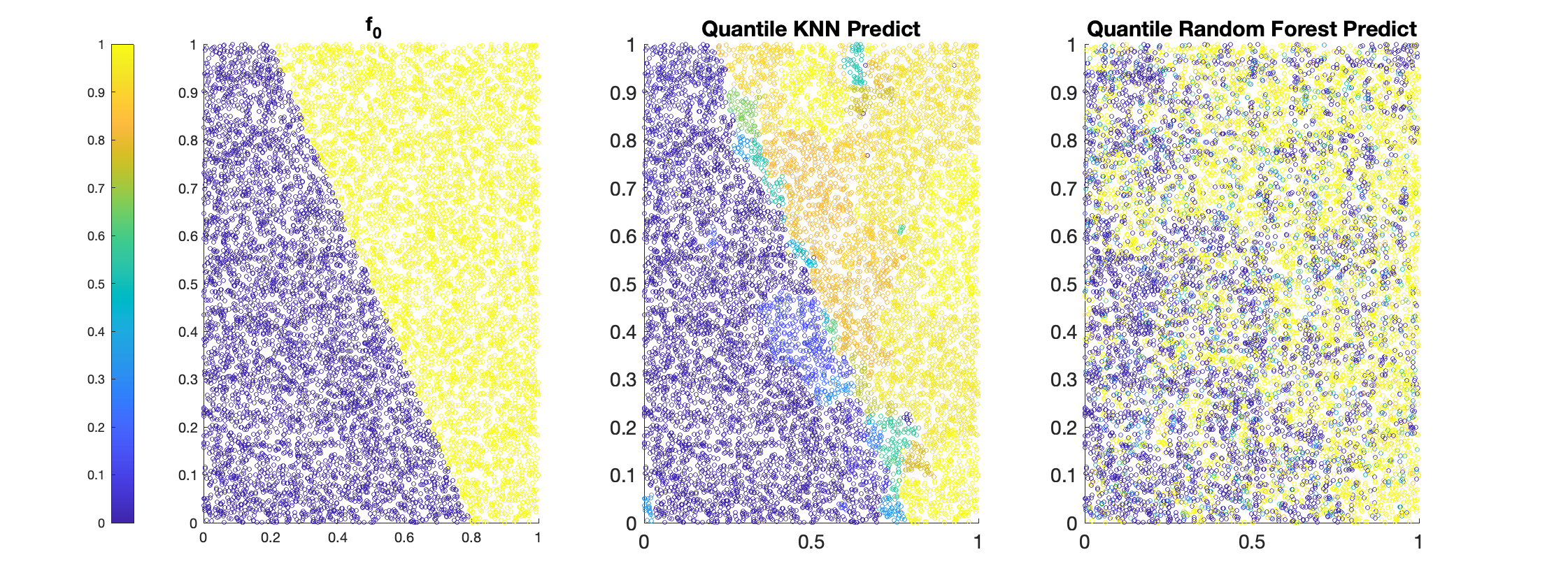}\\
\includegraphics[width = 152mm, height = 44.2mm]{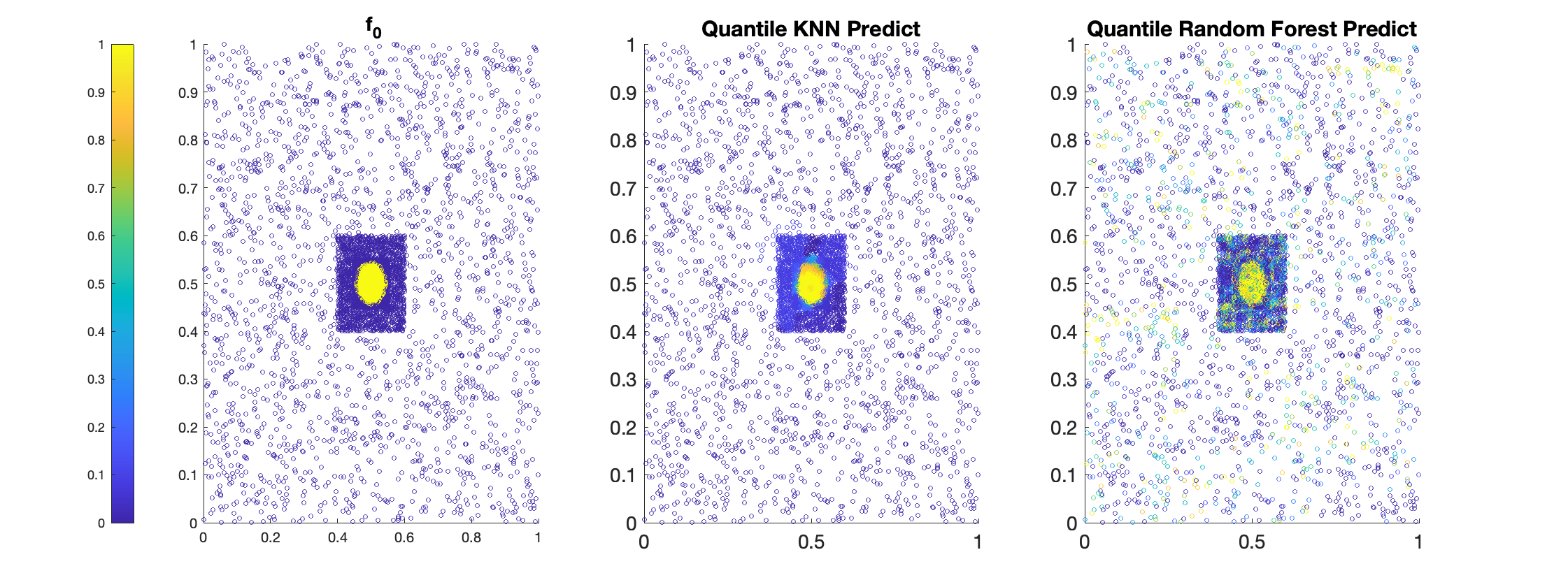}\\
\includegraphics[width = 152mm, height = 44.2mm]{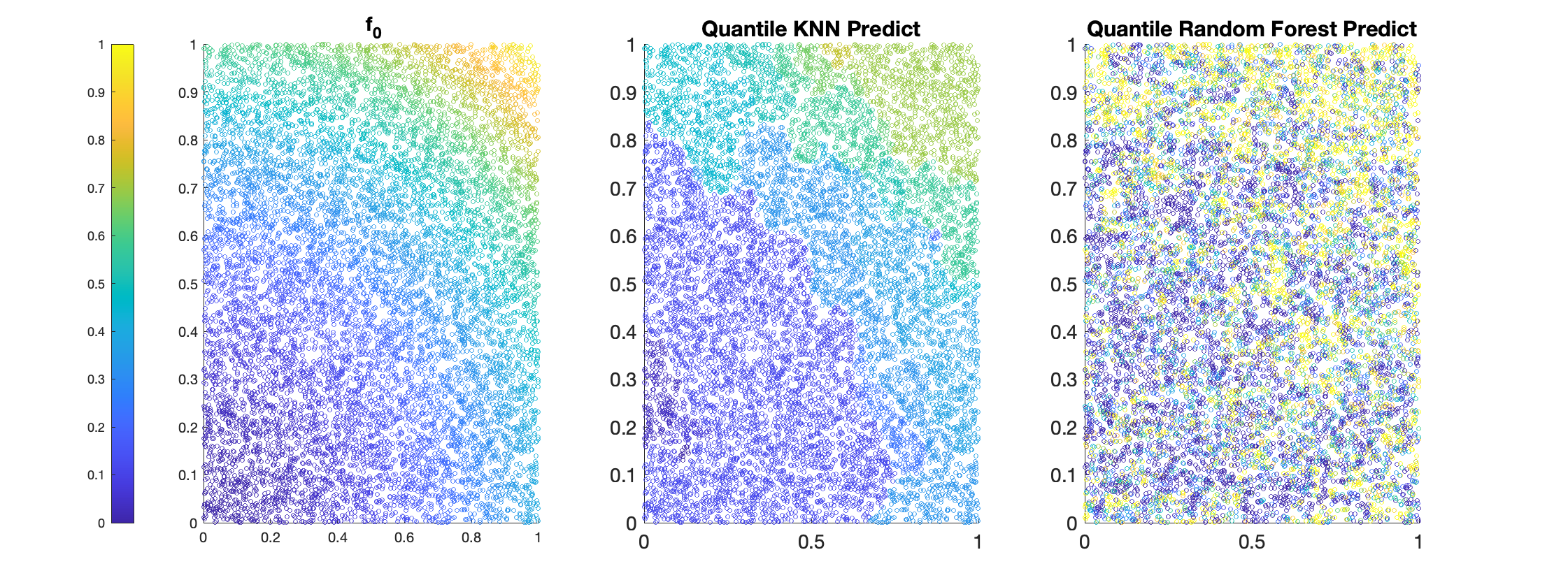}
\caption{Comparison among observations and estimates from Scenario 1 with Gaussian errors (\textit{the first row}), Scenario 1 with Cauchy errors (\textit{the second row}), Scenario 2 (\textit{the third row}), and Scenario 3 (\textit{the fourth row}). \textit{Left column}: 	the function $f_0$  evaluated at the observed $x_i$ for $i =1,\ldots,n$, with $n = 10000$. The horizontal and vertical axis of each panel correspond to the coordinates of  $x_i$.  \textit{Middle  column}: the corresponding estimate of $f_0$ obtained via quantile $K$-NN fused lasso. \textit{Right  column}: the estimate of $f_0$ obtained via quantile random forest.}
\label{fig:1}
\end{figure}

Figure \ref{fig:1} displays the true function and the estimates from both quantile $K$-NN fused lasso and quantile random forest under Scenarios 1, 2 and 3. Clearly, quantile $K$-NN fused lasso provides more reasonable estimation in all cases. Quantile random forest estimates are more noisy and the performance is even poorer under Cauchy errors.

The results presented in Table \ref{table:1} indicate that overall, quantile $K$-NN fused lasso outperforms the competitors in most scenario. As expected, for estimating functions with Gaussian errors like Scenario 1, regular $K$-NN fused lasso is the best method. For Scenarios 2 and 3, when estimating the conditional median of piecewise continuous or continuous functions with heavy-tail errors (such as Cauchy and $t$-distributions), quantile $K$-NN fused lasso achieve the smallest mean square errors over the other two methods.

\begin{figure}[h!]
\centering
\includegraphics[width = 80mm, height = 80mm]{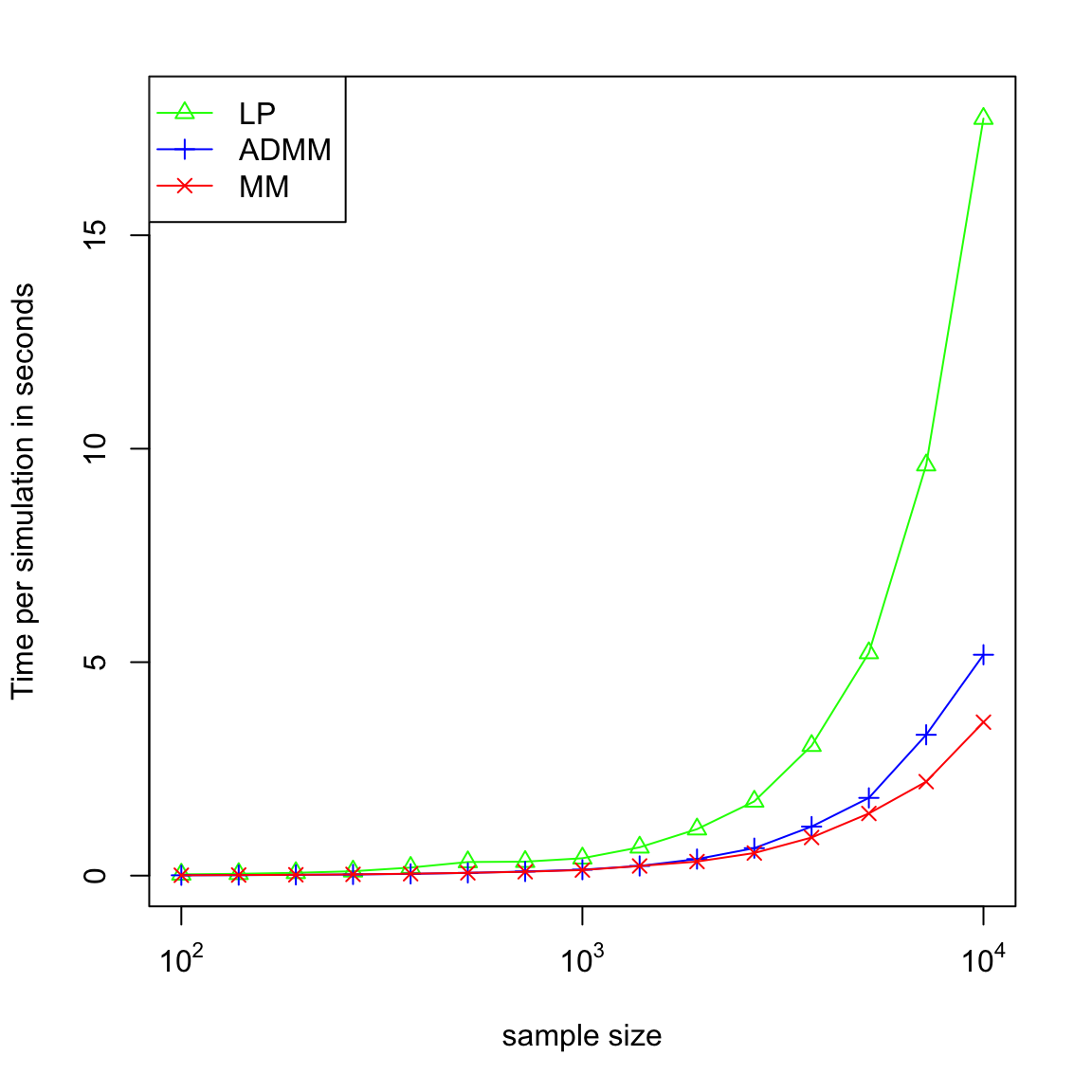}
\caption{A log-scaled plot of time per simulation of LP, ADMM and MM algorithm against problem size $n$ (15 values from $10^2$ up to $10^4$). For each algorithm, the time to compute the estimate for one simulated data is averaged over 100 Monte Carlo simulations.}
\label{fig:2}
\end{figure}

We also compare the performance of linear programming the two algorithms discussed in Section \ref{sec3}, with simulated data from Scenario 3. We obtain almost identical estimators from the three algorithms under the same choice of $\lambda$. Regarding computational time, we record the averaged time consumed over 100 simulations for each algorithm. Figure \ref{fig:2} demonstrates that majorize-minimize (MM) is the most efficient one among the three algorithms, and linear programming (LP) can be very expensive in operational time for large-size problems.

\subsection{Real Data}
\label{sec5.2}
\subsubsection{California Housing Data}
In this section, we conduct an experiment of predicting house value in California based on median income and average occupancy, similar to the experiment in \citet{Petersen16}. The data set, consisting of 20,640 measurements, was originally used in \citet{Pace97} is publicly available from the Carnegie Mellon StatLib data repository (\href{http://lib.stat.cmu.edu}{lib.stat.cmu.edu}).

\begin{figure*}[h!]
	\centering
	\includegraphics[width=0.4\textwidth, height = 46mm]{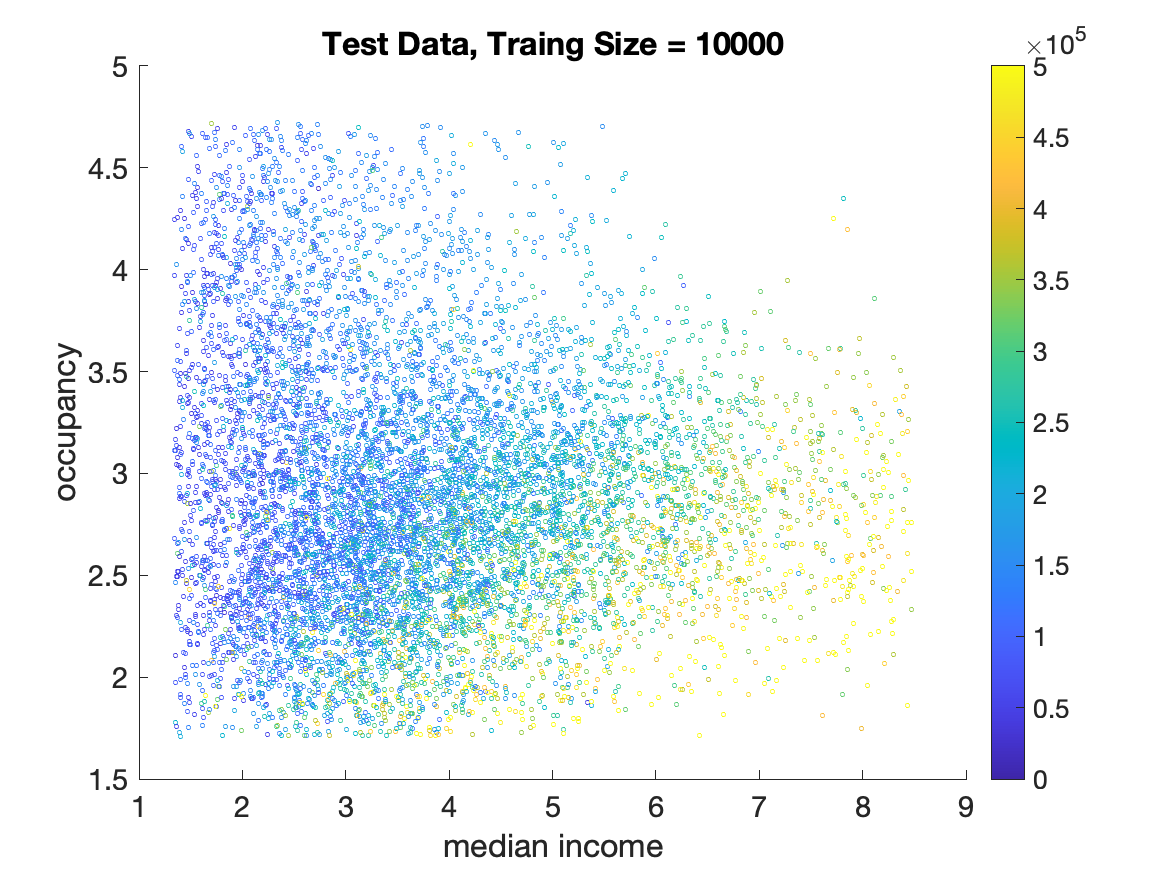} 
	\\
	\includegraphics[width=0.4\textwidth, height = 46mm]{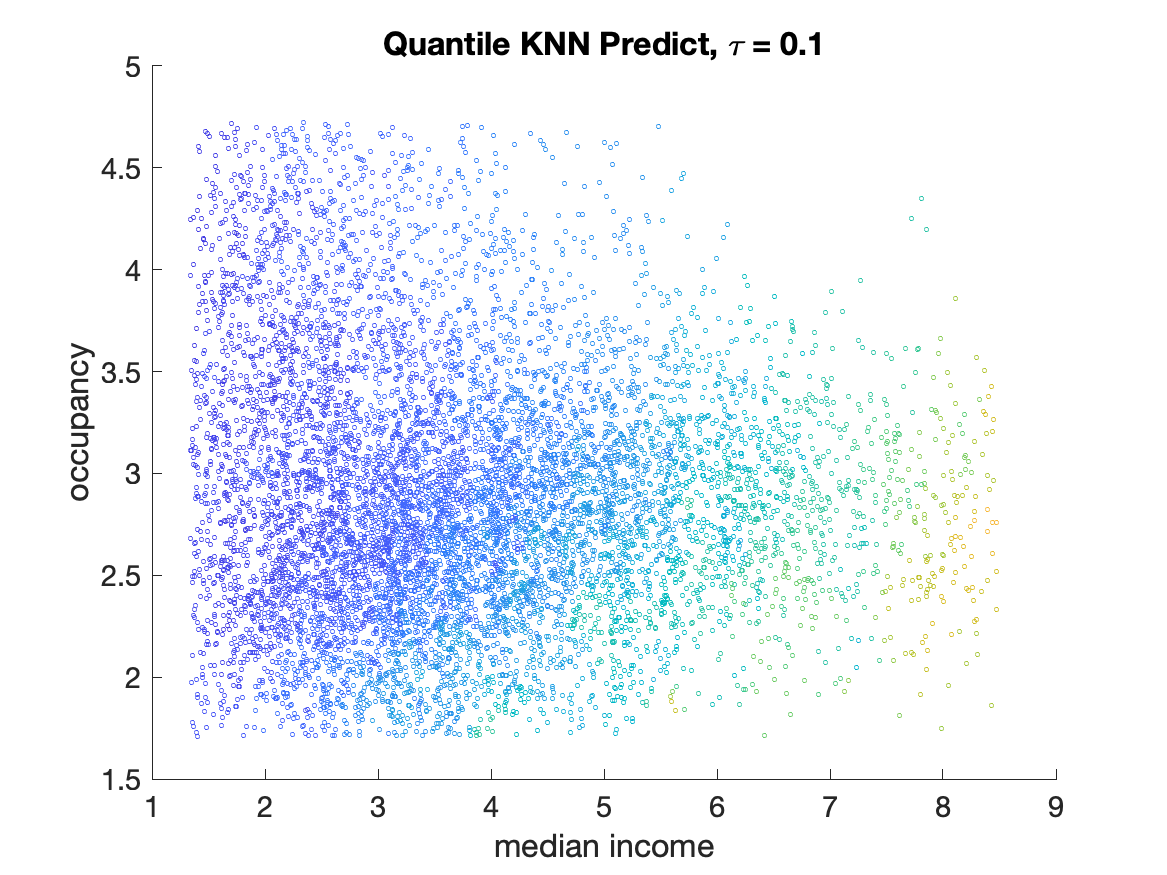}\quad
	\includegraphics[width=0.4\textwidth, height = 46mm]{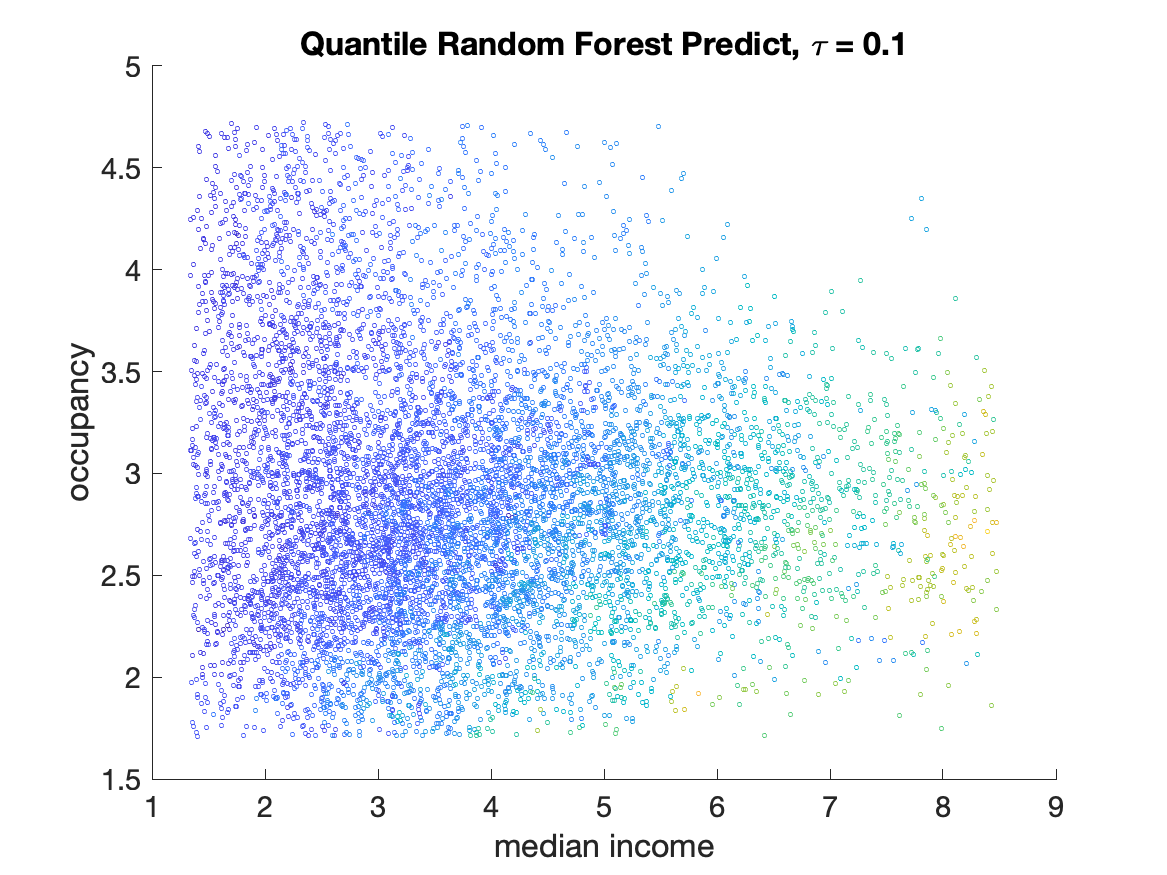}
	\\
	\includegraphics[width=0.4\textwidth, height = 46mm]{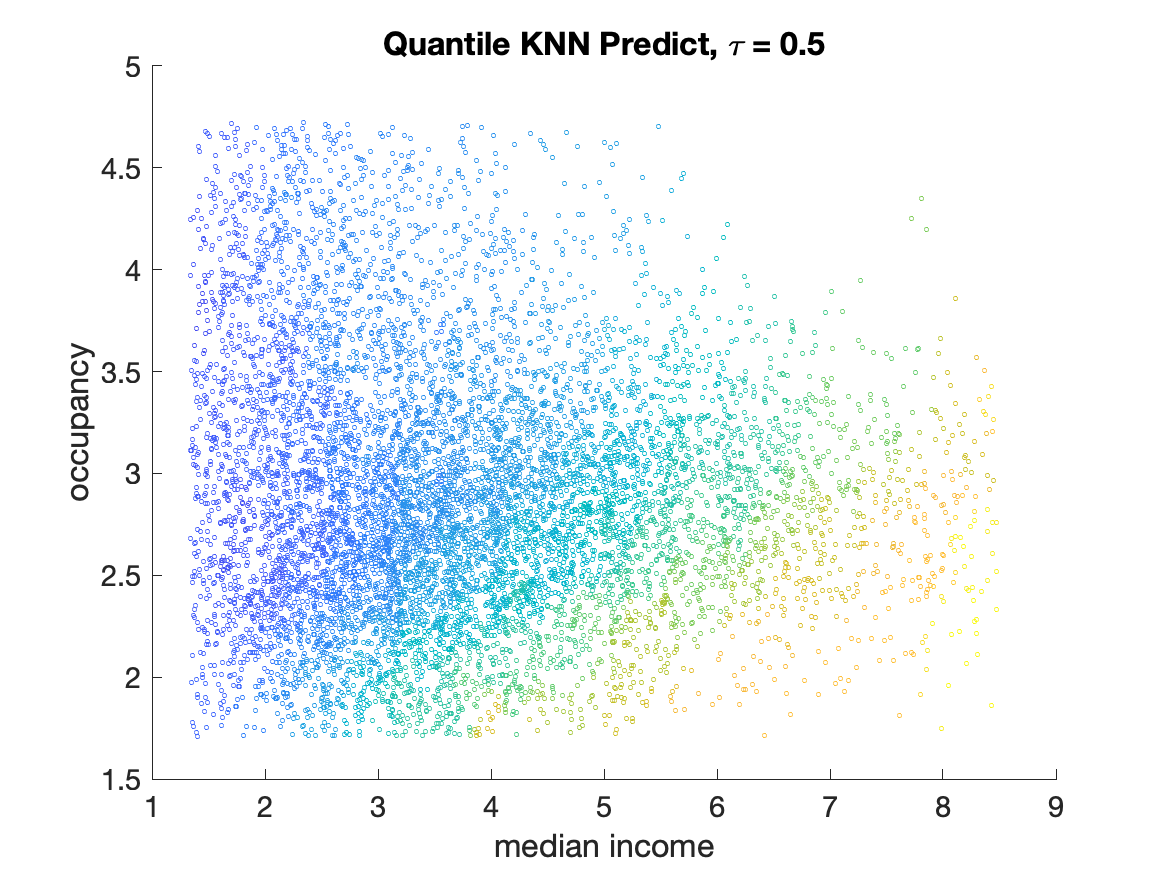}\quad
	\includegraphics[width=0.4\textwidth, height = 46mm]{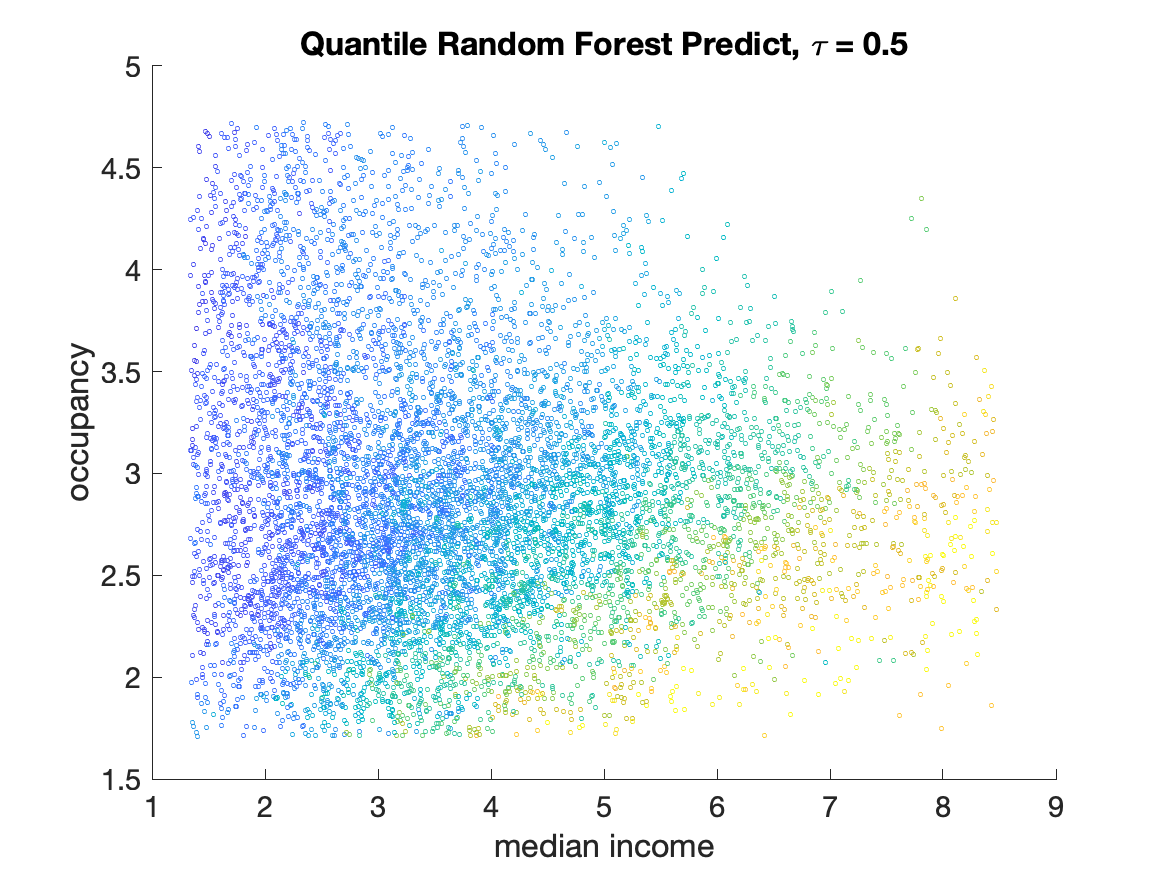}
	\\
	\includegraphics[width=0.4\textwidth, height = 46mm]{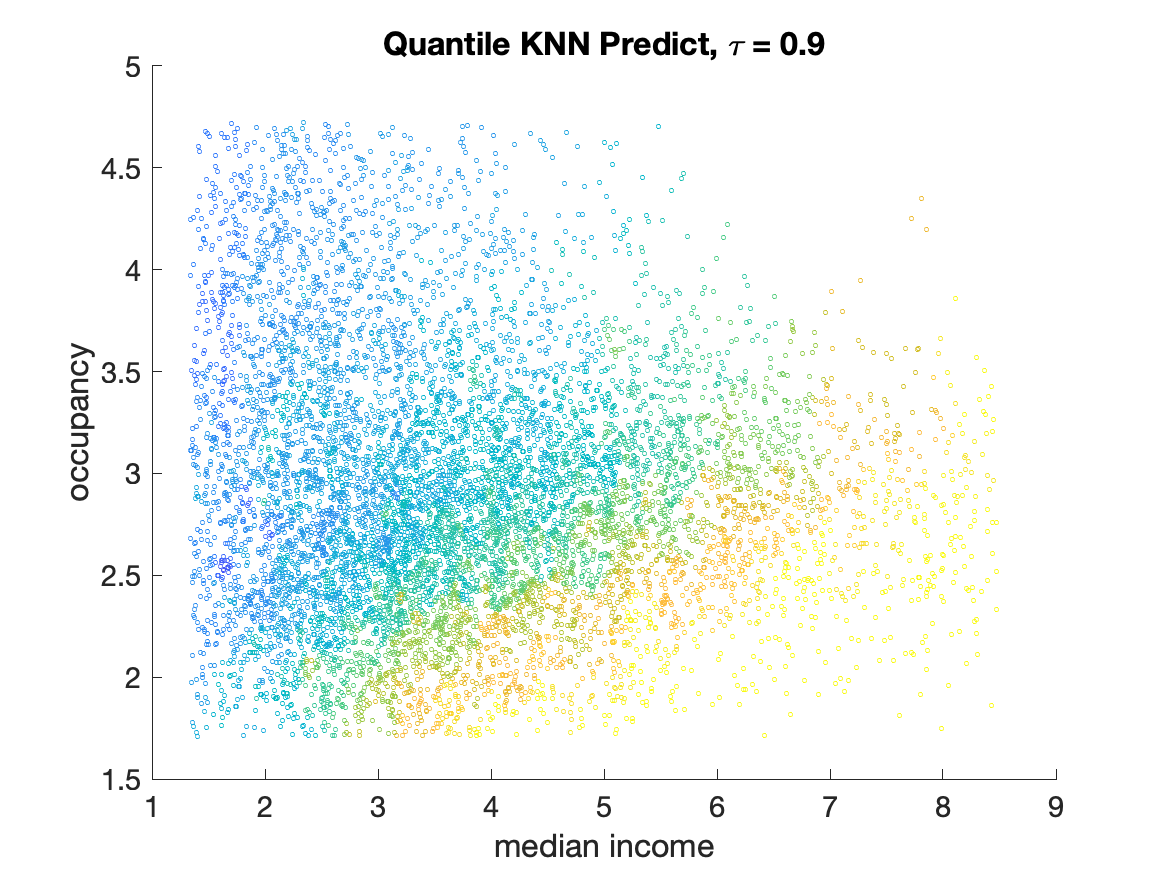}\quad
	\includegraphics[width=0.4\textwidth, height = 46mm]{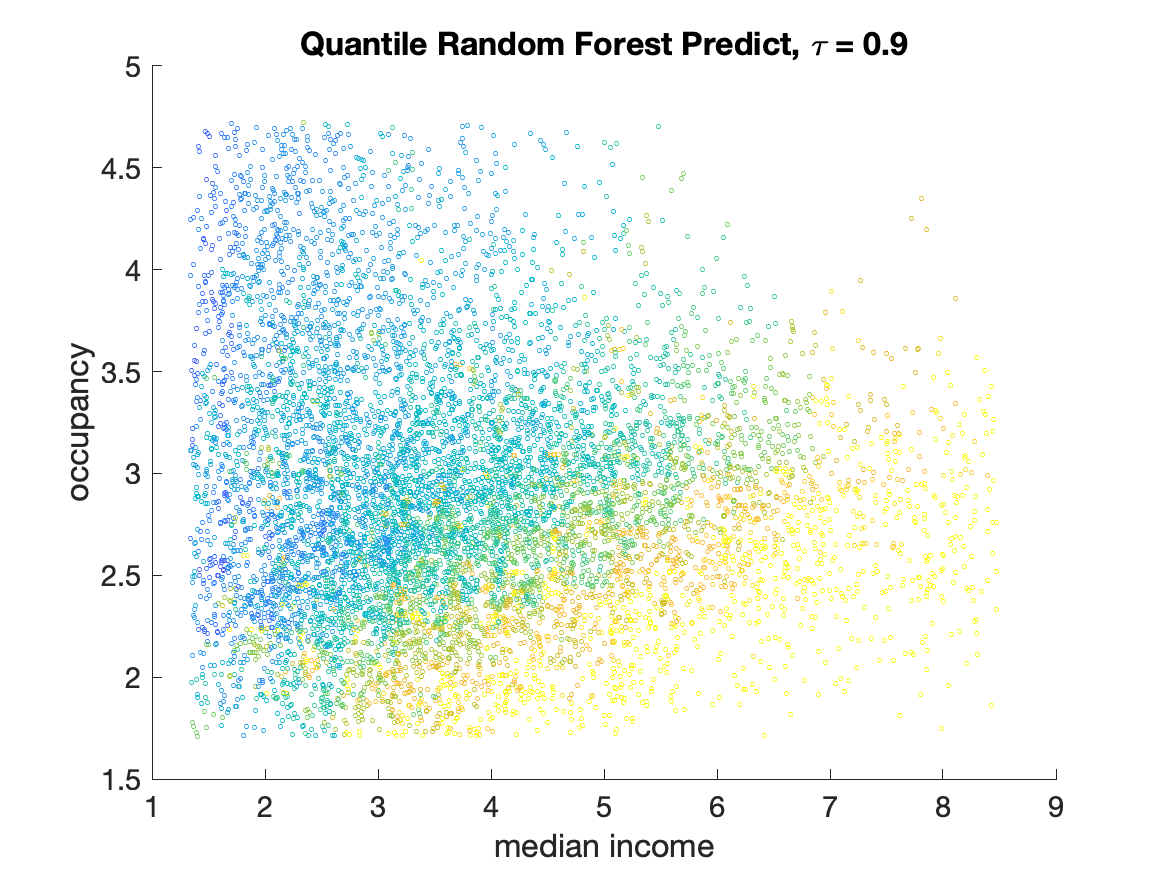}
	\caption{Comparison between predictions from two methods for California housing data, with respect to $\tau = 0.1, 0.5, 0.9$. The plot at the top presents the true testing data value, and the color scale is the same among all plots.}
	\label{fig:3}
\end{figure*}

We perform 100 train-test random splits the data, with training sizes 1000, 5000, and 10000. For each split  the data not in the training set is treated as testing data. For median estimation, we compare averaged mean squared prediction errors  of the test sets  (after taking the log of housing price) from quantile $K$-NN fused lasso and quantile random forest; besides, we construct 90$\%$ and 95$\%$ prediction intervals from both methods and report the  proportion of true observations in the test set that fall in the intervals. Both evaluations are averaged over 100 repetitions for each method. The tuning parameter $\lambda$ for quantile $K$-NN fused lasso is  chosen based on the BIC criteria for each training and the parameters for quantile random forest are selected as default.

From the results in Table \ref{table:2}, quantile $K$-NN fused lasso has better performance than quantile random forest in all cases. The result agrees with the nature of piecewise continuity in housing price, that guarantee the advantage of our proposed method over the competitor. When we illustrate the predicted values  on a test set from  a experiment with  training size of 10,000 visually in Figure \ref{fig:3}, we also observe a piecewise continuous structure in quantile $K$-NN fused lasso estimates, while estimates from quantile random forest contain more noise, especially for lower and higher quantiles.

\begin{table}[h!]
	\centering
	\begin{tabular}{c c c c}
		\hline
		Training Size & $\tau$ & QKNN & QRF\\
		\hline
		1000 & 0.5 & \textbf{0.1553} (0.0032) & 0.1679 (0.0026)\\
		5000 & 0.5 & \textbf{0.1485} (0.0016) & 0.1646 (0.0016) \\
		10000 & 0.5 & \textbf{0.1479} (0.0017) & 0.1623 (0.0019)\\
		\hline
		\\
		\hline
		Training Size & $\tau$& QKNN & QRF \\
		\hline
		1000 & 0.9 & \textbf{0.8405} (0.0141) & 0.8315 (0.0086) \\
		5000 & 0.9& \textbf{0.8569} (0.0076) & 0.8334 (0.0040)\\
		10000 & 0.9 &  \textbf{0.8585} (0.0061) & 0.8337 (0.0041)\\
		1000 & 0.95& \textbf{0.9163} (0.0127) & 0.8887 (0.0084)\\
		5000 &  0.95& \textbf{0.9216} (0.0099) & 0.8891 (0.0040)\\
		10000 &  0.95 &  \textbf{0.9274} (0.0097) & 0.8902 (0.0042) \\
		\hline 
		\\
	\end{tabular}
	\caption{Average test set prediction error (across the 100 test sets) on California housing data. For median, we report the mean squared errors; for other quantiles  ($\tau \in  \{0.9,0.95\}$), we report the averaged proportion of the true data located in the predicted confidence interval. Standard Monte Carlo errors are recorded in parentheses. The number of nearest neighbors, $K$ is chosen as 5 for quantile $K$-NN fused lasso.}
	\label{table:2}
\end{table}

Finally, it is clear from Table \ref{table:2}  that both quantile $K$-NN fused lasso and quantile random forest  have coverage probabilities are noticeably lower than the true nominal level. Since the results in Table \ref{table:2}  correspond to a real data set,  training on a subset and predicting on a different subset of the data, we are not aware of how to improve the coverage for both of the competing methods.

\subsubsection{Chicago Crime Data}
We apply quantile $K$-NN fused lasso and quantile random forest to a dataset of publicly-available crime report counts in Chicago, Illinois in 2015. We preprocess the dataset in the same way as \citet{Tansey18} by merging all observations into a fine-grained 100 $\times$ 100 grid based on latitude and longitude, taking the log of the total counts in each cell, and omitting all cells with zero count. The resulting preprocessed data contains a total number of 3756 data points along the grid. Similar to the experiment on California housing data, we perform a train-test split with training size 500, 1000, 1500, and 2000, and model on the median counts according to the position on the $X$ and $Y$ axes of the grid. We then predict the value on test set and report the averaged square errors over 100 test sets. The parameters for both methods are selected in the same way as in the previous experiment.
\\
\begin{table}[h!]
    \centering
    \begin{tabular}{c c c c c}
        \hline
        Training Size & $\tau$ & QKNN & QRF\\
        \hline
        500 & 0.5 & 1.2161 (0.0441) & \textbf{1.1441} (0.0479)\\
        1000 & 0.5 & 1.0151 (0.0374) & \textbf{0.9836} (0.0417) \\
        1500 & 0.5 & \textbf{0.9138} (0.0343) & 0.8959 (0.0380) \\
        2000 & 0.5 & \textbf{0.8476} (0.0409)& 0.8383 (0.0417) \\
        \hline
        \\
    \end{tabular}
\caption{Average test set prediction errors (across the 100 test sets) with standard errors on Chicago Crime data.}
\label{table:3}
\end{table}

\begin{figure}[h!]
\centering
\includegraphics[width = 152mm, height = 50mm]{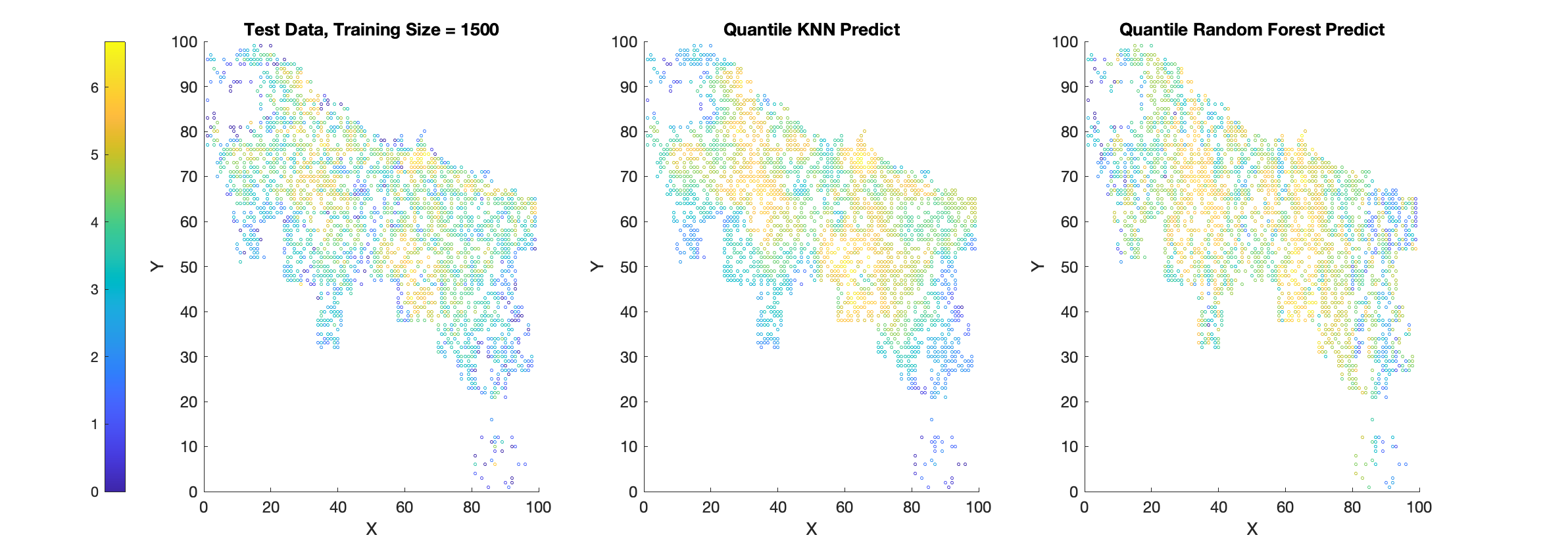}
\caption{\textit{Left}: one test data from Chicago Crime Data, under training size 1500. \textit{Middle}: the estimate obtained via quantile $K$-NN fused lasso. \textit{Right}: the estimate obtained via quantile random forest.}
\label{fig:4}
\end{figure}

From the results of Table \ref{table:3}, we see that quantile $K$-NN fused lasso still outperforms quantile random forest in MSE if the training size is at least 1500. We notice that for small sample size, our method suffers. This is presumably due to the fact that the raw data is not smooth enough. Yet, Figure \ref{fig:4} shows that quantile $K$-NN fused lasso capture local patterns more successfully than quantile random forest in most regions.

\section{Conclusion}
\label{sec6}

In this paper we have proposed and studied a method  for estimating of quantile functions that are piecewise constant. Such classes of functions naturally arise in  real-life data. Estimating accurately signals with such structure is an important but challenging research topic arising in many practical contexts. However, there is a lack of existing literature on robust estimation of piecewise constant or piecewise Lipschitz signals in multivariate settings. Motivated by this, we have proposed the quantile $K$-NN fused lasso estimator.

Our numerical  experiments on different real-life and simulated examples  lay out empirical evidence of the superior robustness  and local adaptivity of our estimator over other state-of-the-art methods. We have presented two algorithms to efficiently compute the solutions to the corresponding optimization problem behind our method. Our theoretical analysis extends previous work on $K$-NN fused lasso \citep{Padilla20a} and quantile trend filtering \citep{Madrid20}. Specifically, we have shown  that the quantile $K$-NN fused lasso estimator achieves an optimal convergence rate of $n^{-1/d}$ for estimating a  $d$-dimensional piecewise Lipschitz quantile function. This result is guaranteed under mild assumptions on data, and thus our estimator can be applied to more general models rather than  only those with sub-Gaussian errors.



\appendix
\section*{Appendix A. Closed-form Solution to the Primal in ADMM Algorithm}
\label{secA}
In Section \ref{sec3.1}, we introduce an ADMM algorithm to compute quantile $K$-NN estimates. The algorithm requires to solve the primal problem (\ref{eq12})
\begin{equation*}
    \ \ \ \ \theta \leftarrow \underset{\theta\in \mathbb{R}^n}{\arg\min} \ \left\{\sum_{i=1}^n\rho_\tau(y_i-\theta_i) + \frac{R}{2}\Vert\theta - z + u\Vert^2\right\}.
\end{equation*}
We can solve the problem coordinate-wisely: for $i = 1,...,n$, we find the minimizer 
\begin{equation*}
    \ \ \ \ \theta_i \leftarrow \underset{\theta_i \in \mathbb{R}}{\arg\min} \ \left\{\rho_\tau(y_i-\theta_i) + \frac{R}{2}(\theta_i - z_i + u_i)^2\right\}.
\end{equation*}
By definition, 
$$\rho_\tau(y_i-\theta_i) = \begin{cases}
\tau(y_i-\theta_i) & \text{ if } \  y_i - \theta_i > 0, \\
(\tau-1)(y_i-\theta_i) & \text{ if } \  y_i - \theta_i < 0, \\
0 & \text{ if } \ y_i - \theta_i = 0.
\end{cases}$$
We discuss the three cases separately. \\ \\
(1) When $y_i - \theta_i > 0$, $\theta_i \leftarrow \arg\min \left\{\tau(y_i-\theta_i) + \frac{R}{2}(\theta_i - z_i + u_i)^2\right\}$. Take the derivative and set to 0 to obtain $\theta_i = z_i - u_i + \tau/R$. The condition $y_i - \theta_i > 0$ then becomes $y_i - z_i + u_i > \tau/R$. \\ \\
(2) When $y_i - \theta_i < 0$, $\theta_i \leftarrow \arg\min \left\{(\tau-1)(y_i-\theta_i) + \frac{R}{2}(\theta_i - z_i + u_i)^2\right\}$. Take the derivative and set to 0 to obtain $\theta_i = z_i - u_i + (\tau-1)/R$. The condition $y_i - \theta_i < 0$ then becomes $y_i - z_i + u_i < (\tau-1)/R$. \\ \\
(3) When $y_i - \theta_i = 0$, it is simple to get $\theta_i = y_i$. \\ \\
To summarize, the closed-form solution to the primal (\ref{eq12}) is
\begin{equation*}
\theta_i = \begin{cases}
z_i - u_i + \frac{\tau}{R} & \text{ if } y_i - z_i + u_i > \frac{\tau}{R},\\
z_i - u_i + \frac{\tau-1}{R} & \text{ if } y_i - z_i+ u_i < \frac{\tau-1}{R},\\
y_i & \text{ otherwise; }
\end{cases}
\end{equation*}
for $i = 1,...,n$.

\section*{Appendix B. Sensitivity of Initialization in the ADMM Algorithm}

In this section, we examine how the convergence  of  Algorithm \ref{alg1} is sensitive to different initials. Recall that Algorithm \ref{alg1} 
 requires an initialization of $\theta^{(0)}$, $z^{(0)}$ and $u^{(0)}$. A practical way to choose these initials is to set $\theta^{(0)} = z^{(0)} = y, u^{(0)} = 0$, where $y$ is the input data, as defined in Algorithm \ref{alg1}. Here, we try three other initializations to assess the sensitivity to the initials: \\ \\
(1) The setup in Algorithm \ref{alg1}: $\theta^{(0)} = z^{(0)} = y, u^{(0)} = 0$. \\ \\
(2) We add perturbations to the first setup: $\theta^{(0)} = y + v_1, z^{(0)} = y + v_2, u^{(0)} = v_3$, where $v_1,v_2,v_3  \overset{\text{ind}}{\sim}  N(0,I_n)$.  \\ \\
(3) Random initialization: $\theta^{(0)} = w_1, z^{(0)} = w_2, u^{(0)} = w_3$, where $w_1,w_2,w_3 \overset{\text{ind}}{\sim}  N(0,I_n)$. \\ \\
(4) Random initialization with a large multiplication factor: $\theta^{(0)} = 50s_1, z^{(0)} = 50 s_2, u^{(0)} = 50s_3$, where $s_1,s_2,s_3 \overset{\text{ind}}{\sim}  N(0,I_n)$. \\

We use the data generation mechanism in Scenario 2 from the main content, and errors are independently drawn from a  $t$-distribution with 3 degrees of freedom. The regularization parameter $\lambda$ is set to be fixed as 0.5. We replicate the simulation 500 times with a sample size $n=10000$. Note that the input data and perturbations are generated independently in every replication. For each initialization setup, we record the averaged value of the objective function $\sum_{i=1}^n \rho_\tau(y_i-\theta_i) + \lambda \Vert \nabla_{G} \theta\Vert_1$ after each iteration and the number of iterations to converge when we set $\kappa =0.01$ in Step 3 of Algorithm \ref{alg1}.

\begin{figure}[h!]
\centering
\includegraphics[width = 110mm, height = 85mm]{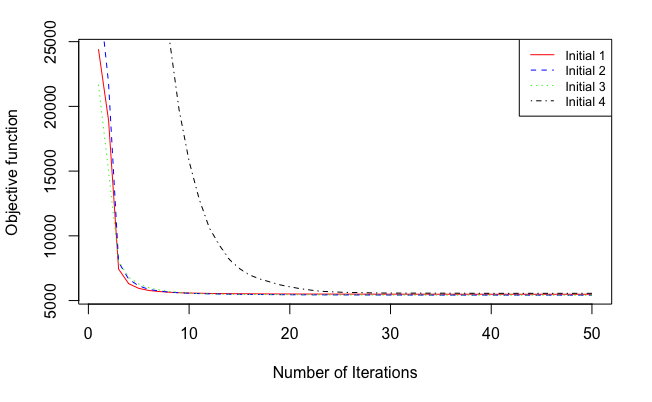}
\caption{Averaged value of objective function (over 500 replications) after each iteration step for different initializations.}
\label{fig:5}
\end{figure}

From Figure 5, we observe that the value of the objective function converges very fast to the optimal minimum within only tens of iterations, regardless of the random initialization that is used. If we look at the number of iterations shown in Table 4, the first three initializations have relatively the same convergence speed. If there exists a large deviation between the initial value and the true signal like in the last setup, the ADMM algorithm takes slightly longer time to converge. Overall, the ADMM algorithm is insensitive to random initialization and fast convergence is observed in practice.\\

\begin{table*}[h!]
    \centering
        \begin{tabular}{c c}
        \hline
        Initialization  & Averaged number of iterations to converge \\
        \hline
        (1) & 11 \\
         (2) & 11.13 \\
         (3) & 11 \\
         (4) & 26.65 \\
         \hline
    \end{tabular}
    \caption{Number of iterations to converge under different initializations, averaging over 500 Monte Carlo simulations.}
    \label{table:4}
\end{table*}

\section*{Appendix C. General Lemmas}
\label{secB}

The following lemmas hold by conditioning on $\{x_i\}_{i=1}^n$.

\begin{definition}
\label{def1}
The function $\Delta^2$ is defined as 
$$\Delta^2(\delta):=\sum_{i=1}^n \min\{|\delta_i|, \delta_i^2\},$$
where $\delta_i \in \mathbb{R}^n$. We also write $\Delta(\delta) = \{\Delta^2(\delta)\}^{1/2}$.
\end{definition}

\begin{definition}
\label{def2}
For a set $S \subset \mathbb{R}^n$, the sub-Gaussian width of $S$ is defined as
$$SGW(S) = \mathbb{E}\left(\sup_{v\in S} \sum_{i=1}^n s_iv_i\right),$$
where $s_1, ...,s_n$ are independent 1-subgaussian random variables.
\end{definition}
The notation of sub-Gaussian width is not used very often in literature compared to a similar definition of Gaussian width, 
$$GW(S) = \mathbb{E}\left(\sup_{v\in K} \sum_{i=1}^n z_iv_i\right),$$
where $z_1, ...,z_n$ are independent standard normal random variables. In fact, the sub-Gaussian width shares many common properties with the Gaussian width, as we can upper bound the sub-Gaussian width by a constant times the Gaussian width using generic chaining; see Chapter 4 in \citet{Talagrand05} and also \citet{Banerjee14} for precise explanations. \\
\begin{definition}
\label{def3}
We define the empirical loss function 
$$\hat M(\theta) = \sum_{i=1}^n \hat M_i(\theta_i),$$
where 
$$\hat M_i(\theta_i) = \rho_\tau(y_i-\theta_i) - \rho_\tau(y_i-\theta^*_i).$$
\end{definition}
Setting $M_i(\theta_i) = \mathbb{E}(\rho_\tau(y_i-\theta_i) - \rho_\tau(y_i-\theta^*_i))$, the population version of $\hat M$ becomes 
$$M(\theta) = \sum_{i=1}^n M_i(\theta_i).$$
Now, we consider the $M$-estimator
\begin{equation}
\label{eq17}
\begin{split}
\hat \theta \ \ = \ \ &\underset{\theta \in \mathbb{R}^n}{\arg\min} \ \ \ \ \  \hat M(\theta) \\
    &\text{subject to} \ \ \theta \in S,
\end{split}
\end{equation}
and $\theta^* \in \arg\min_{\theta \in \mathbb{R}^n} M(\theta)$. Throughout, we assume that $\theta^* \in S \subset \mathbb{R}^n$. \\

\begin{lemma}
\label{lemma1}
With the notation from before,
$$M(\hat\theta) \leq \sup_{v \in S} \left\{M(v) - \hat M(v)\right\}.$$
\end{lemma}
\noindent \textit{Proof.} See the proof of Lemma 10 in \citet{Madrid20}. \\

\begin{lemma}
\label{lemma2}
Suppose that Assumption \ref{as2} holds. Then there exists a constant $c_\tau > 0$ such that for all $\delta \in \mathbb{R}^n$, we have
$$M(\theta+\delta) \geq c_\tau \Delta^2(\delta).$$
\end{lemma}
\noindent \textit{Proof.} See the proof of Lemma 14 in \citet{Madrid20}. \\

\begin{corollary}
\label{corollary3}
Under Assumption \ref{as2}, if $\theta^* \in S$, we have that
\begin{equation}
\label{eq18}
    \mathbb{E}\left\{\Delta^2(\hat \theta - \theta^*)\right\} \leq \mathbb{E}\left[\sup_{v\in S} \left\{M(v)-\hat M(v)\right\}\right].
\end{equation}
\end{corollary}
Next, we proceed to bound the right hand side of (\ref{eq18}).
\\
\begin{lemma}
\label{lemma4}
(Symmetrization) Under Assumption \ref{as2}, we have that
$$\mathbb{E}\left[\sup_{v\in S} \left\{M(v)-\hat M(v)\right\}\right] \leq 2\mathbb{E}\left\{\sup_{v\in S} \sum_{i=1}^n \xi_i\hat M_i(v_i)\right\},$$
where $\xi_1,...,\xi_n$ are independent Rademacher variables independent of $y_1,...,y_n$.
\end{lemma}
\noindent \textit{Proof.} See the proof of Lemma 11 in \citet{Madrid20}.  \\

\begin{lemma}
\label{lemma5}
(Contraction principle) Under Assumption \ref{as2}, we have that
$$\mathbb{E}\left\{\sup_{v\in S} \sum_{i=1}^n s_i\hat M_i(v_i)\right\}\leq 2SGW(S-\theta^*) = 2SGW(S),$$
where $s_1, ...,s_n$ are independent 1-subgaussian random variables independent of $y_1,...,y_n$.
\end{lemma}
\noindent \textit{Proof.} Recall that $\hat M_i(v_i) =\rho_\tau(y_i-v_i) - \rho_\tau(y_i-\theta^*_i)$. Clearly, these are 1-Lipschitz continuous functions. Therefore,
\begin{align*}
    \mathbb{E}\left\{\sup_{v\in S} \sum_{i=1}^n s_i\hat M_i(v_i)\right\} & = \mathbb{E}\left(\mathbb{E}\left\{\sup_{v\in S} \sum_{i=1}^n s_i\hat M_i(v_i) \bigg| y\right\}\right) \\
    &\leq \mathbb{E}\left(\mathbb{E}\left\{\sup_{v\in S} \sum_{i=1}^n s_iv_i \bigg| y\right\}\right) \\
    &= \mathbb{E}\left\{\sup_{v\in S} \sum_{i=1}^n s_i(v_i-\theta^*)\right\} + \mathbb{E}\left(\sum_{i=1}^n s_i\theta^*\right) \\
    &= \mathbb{E}\left\{\sup_{v\in S} \sum_{i=1}^n s_i(v_i-\theta^*)\right\}
\end{align*}
where the inequality follows from the Gaussian version of Talagrand’s contraction principle, see corallary 3.17 in \citet{Ledoux13} and 7.2.13 in \citet{Vershynin18}. $\hfill{\square}$

\section*{Appendix D. $K$-NN Embeddings}
\label{secC}
This section follows from Section D and E in \citet{Padilla20a} and it originally appeared in the flow-based proof of Theorem 4 in \citet{vonLuxburg14}. The main idea is to embed a mesh on a $K$-NN graph corresponding to the given observations $X = (x_1,...,x_n)$ under Assumptions \ref{as3}-\ref{as5} so that we are able to bound the total variation along the grid graph from below. In this way, we can further derive an upper bound on the loss of the optimization problem (\ref{eq8}) and (\ref{eq10}) with respect to the loss function defined in Definition \ref{def1}. 

First, we need to construct a valid grid discussed in Definition 17 of \citet{vonLuxburg14} with a minor modification. With high probability, a valid grid graph $G$ satisfies the following: (i) the grid width is not too small: each cell of the grid contains at least one of the design points; (ii) the grid width is not too large: points in the same or neighboring cells of the grid are always connected in the $K$-NN graph. 

Given $N \in \mathbb{N}$, a $d$-dimensional grid graph $G_\text{lat} \in [0,1]^d$ has vertex set $V_\text{lat}$ and edge set $E_\text{lat}$. The grid graph has equal side lengths, and the total number of nodes $|V_\text{lat}|=N^d$. Without loss of generality, we assume that the nodes of the grid correspond to the points
\begin{equation}
\label{eq19}
    P_\text{lat}(N) = \left\{\left(\frac{i_1}{N}- \frac{1}{2N},...,\frac{i_d}{N}- \frac{1}{2N}\right):i_1,...,i_d \in \{1,...,N\}\right\}.
\end{equation}
Moreover, for $z, z' \in P_\text{lat}(N)$, $(z,z') \in E_\text{lat}(N)$ if and only if $\Vert z-z'\Vert_2 = \frac{1}{N}$. 

Now, we define $I(N) = h^{-1}\{P_\text{lat}(N)\}$ as the mesh in the covariate space $\mathcal{X}$ corresponding to the grid graph $G_\text{lat}(N) \in [0,1]^d$ through the homeomorphism $h$ from Assumption \ref{as5}. In general, $I(N)$ performs as a quantifization in the domain $\mathcal{X}$; see \citet{Alamgir14} for more details. We denote the elements in $I(N)$ by $u_1,...,u_{N^d}$, and define a collection of cells $\{C(x)\} \in \mathcal{X}$ for $x \in I(N)$ as 
\begin{equation}
\label{eq20}
    C(x) = h^{-1}\left(\left\{z \in [0,1]^d:h(x) = \underset{x'\in P_\text{lat}(N)}{\arg\min} \ \Vert z-x'\Vert_\infty\right\}\right).
\end{equation}
In order to analyze the behavior of the proposed estimator $\hat\theta$ through the grid embedding, we construct two vectors denoted by $\theta_I \in \mathbb{R}^n$ and $\theta^I \in \mathbb{R}^{N^d}$ for any signal $\theta \in \mathbb{R}^n$. 

The first vector, $\theta_I \in \mathbb{R}^n$  incorporates information about the samples $X = (x_1,...,x_n)$ and the cells $\{C(x)\}$. The idea is to force covariates $x_i$ fallen in the same cell to take the same signal value after mapping with the homeomorphism. Formally, we define
\begin{equation}
\label{eq21}
    (\theta_I)_i = \theta_j \ \ \ \ \text{where } \ j = \underset{l = 1,...,n}{\arg\min} \ \Vert h(P_I(x_i))-h(x_l)\Vert_\infty,
\end{equation}
where $P_I(x)$ is the point in $I(N)$ such that $x \in C(P_I(x))$ and if there exists multiple points satisfying the condition, we arbitrarily select one.

The second vector, $\theta^I \in \mathbb{R}^{N^d}$ records coordinates corresponding to the different nodes of the mesh (centers of cells), and is associated with $\theta_I$. We first induce a signal in $\mathbb{R}^{N^d}$ corresponding to the elements in $I(N)$ as
$$I_j = \{i = 1,...,n : P_I(x_i) = u_j\}, \ \text{for } j = 1,...,N^d.$$
If $I_j \neq \emptyset$, then there exists $i_j \in I_j$ such that $(\theta_I)_i = \theta_{i_j}$ for all $i \in I_j$; if $I_j$ is empty, we require $\theta_{i_j} = 0$.
We can thus define $$\theta^I = \left(\theta_{i_1},..., \theta_{i_{N^d}}\right).$$
Note that in the proof of Lemma 11 in \citet{Padilla20a}, the authors showed that under Assumptions \ref{as3}-\ref{as5}, with probability approaching 1, 
\begin{equation}
\label{eq22}
    \max_{x\in I(N)} |C(x)| \leq \textup{poly}(\log n).
\end{equation}
We will use this inequality in our proofs later, but we will not make the polynomial function of $\log n$ explicit.

\section*{Appendix E. Definition of Piecewise Lipschitz}
\label{secD}
To make sure Assumption \ref{as1} is valid, we require a piecewise Lipschitz condition on the regression function $f_0$. In this section, we provide the detailed definition of the class of piecewise Lipschitz functions, followed from Definition 1 in \citet{Padilla20a}. All notations follow the same from the main context, besides an extra notation on the boundary of a set $A$, denoted by $\partial A$.

\begin{definition}
	\label{def_lip}
Let $\Omega_\epsilon:= [0,1]^d \backslash B_\epsilon(\partial[0,1]^d)$. We say that a bounded function $g:[0,1]^d\rightarrow\mathbb{R}$ is \textit{piecewise Lipschitz} if there exists a set $\mathcal{S}\subset (0,1)^d$ that has the following properties:
\begin{itemize}
  \item The set $\mathcal{S}$ has Lebesgue measure zero.
  \item For some constants $C_{\mathcal{S}}, \epsilon_0 > 0$, we have that $\mu(h^{-1}\{B_\epsilon(\mathcal{S}) \cap ([0,1]^d\backslash\Omega_\epsilon)\}) \leq C_{\mathcal{S}} \epsilon$ for all $0 < \epsilon < \epsilon_0$.
  \item There exists a positive constant $L_0$ such that if $z$ and $z'$ belong to the same connected component of $\Omega_\epsilon\backslash B_\epsilon(\mathcal{S})$, then $|g(z)-g(z')|\leq L_0\Vert z-z'\Vert_2$.
\end{itemize}
\end{definition}
Roughly speaking, a bounded function $g$ is piecewise Lipschitz if there exists a small set $\mathcal{S}$ that partitions $[0,1]^d$ in such a way that $g$ is Lipschitz within each connected component of the partition. Theorem 2.2.1 in \citet{Ziemer12}  implies that if $g$ is piecewise Lipschitz, then $g$ has bounded variation on any open set within a connected component.

\section*{Appendix F. Theorem \ref{theorem1}}
\label{secE}
\subsection*{F.1 \ \ Notations}
For a matrix $D$, we denote its kernel by $\text{Ker}(D)$ and its Moore–Penrose inverse by $D^\dagger$. We also write $\Pi$ as the projection matrix onto $\text{Ker}(D)$, and denote $\text{Ker}(D)^\perp$ as the orthogonal complement to the kernel space of $D$. 

Our goal is to upper bound the expectation of $M(\cdot) -\hat M(\cdot)$ in the constrained set $S$ defined by
\begin{equation}
\label{eq23}
    S = \left\{\theta \in \mathbb{R}^n \ :  \ \Vert \nabla_G \theta\Vert_1 \leq V n^{1-1/d}\textup{poly}(\log n)\right\},
\end{equation}
where $V = \Vert \nabla_G\theta^*\Vert_1/[n^{1-1/d}\textup{poly}(\log n)]$ and $V\geq V^*$. Through the $K$-NN embedding, we can instead bound the expected loss in the embedded set defined by 
\begin{equation}
\label{}
    \tilde{S} = \left\{\theta \in \mathbb{R}^n \ :  \ \Vert D \theta^I\Vert_1 \leq V n^{1-1/d}\textup{poly}(\log n)\right\},
\end{equation}
where $\theta^I$ follows the definition in Appendix D.

\subsection*{F.2 \ \ Auxiliary lemmas for Proof of Theorem \ref{theorem1}}
\begin{lemma}
\label{lemma6}
For $v \in \mathbb{R}^n$, if $\Vert \nabla_G v\Vert_1 \leq \tilde V$, and $\Delta^2(v-\theta^*) \leq t^2$, then $\Vert D v^I\Vert_1 \leq \tilde V$, and $\Delta^2(v^I - \theta^{*,I}) \leq c_1t^2$ for some constant $c_1$.
\end{lemma}
\noindent \textit{Proof.} 
Lemma 4 in \citet{Padilla20a} obtains the inequality
$$\Vert D v^I\Vert_1 \leq \Vert \nabla_G v\Vert_1, \ \forall v \in \mathbb{R}^.$$
Hence, the first claim follows. \\ \\
Next, we observe that for any vector $u \in \mathbb{R}^n$, 
$$\Delta^2(u^I) =  \left(\sum_{j=1}^{N^d}\min\left\{|u_{i_j}|, u_{i_j}^2\right\}\right) \leq \left(\sum_{i=1}^{n}\min\left\{|u_i|, u_i^2\right\}\right) \leq \Delta^2(u),$$
for some positive constant $c_1$. The second claim follows then. $\hfill \square$
\\ \\
Lemma \ref{lemma6} gives us the fact that 
\begin{equation}
\label{}
\left\{v\in S:\Delta^2(v-\theta^*)\leq t^2\right\} \subseteq \left\{v\in \tilde S:\Delta^2(v^I-\theta^{*,I})\leq c_1t^2\right\}.
\end{equation}
\begin{lemma}
\label{lemma7} Under Assumptions \ref{as1}-\ref{as5}, we have that
$$ \begin{array}{lll}
\displaystyle \mathbb{E}\left[\sup_{v \in \tilde S:\Delta^2(v^I-\theta^{*,I})\leq t^2}\sum_{i=1}^n\xi_i(v_i-\theta_i^*)\right] &\leq &\displaystyle \textup{poly}(\log n)\mathbb{E}\left[\sup_{v \in \tilde S:\Delta^2(v^I-\theta^{*,I})\leq t^2}\sum_{j=1}^{N^d}\tilde{\xi}_j(v^I_j-\theta^{*,I}_j)\right]  \\
&&\displaystyle +  2 V n^{1-1/d}\textup{poly}(\log n),
\end{array} $$
where $\tilde \xi \in \mathbb{R}^{N^d}$ is a 1-subgaussian vector whose coordinates are independent.
\end{lemma}
\noindent \textit{Proof.} We notice that
\begin{align*}
    \xi^\top(v - \theta^*) &= \xi^\top(v - v_I) + \xi^\top(v_I - \theta_I^*) + \xi^\top(\theta_I^* - \theta^*) \\
    &\leq 2\|\xi\|_{\infty}  \left(\Vert\nabla_G v\Vert_1 + \Vert\nabla_G \theta^*\Vert_1\right) + \xi^\top(v_I - \theta_I^*),
\end{align*}
where the second inequality holds by Lemma 4 in \cite{Padilla20a}.
Moreover, 
$$\xi^\top(v_I - \theta_I^*) = \sum_{j=1}^{N^d}\sum_{l\in I_j}\xi_l (v_{i_j} -\theta^*_{i_j}) = \left\{\max_{u\in I}|C(u)|\right\}^{1/2} \tilde \xi^\top(v^I -\theta^{*,I})$$
where
$$\tilde{\xi}_j = \left\{\max_{u\in I}|C(u)|\right\}^{-1/2}\sum_{l \in I_j} \xi_l.$$
Clearly, the $\tilde \xi_1,...,\tilde \xi_{N^d}$ are independent and also 1-subgaussian as the original Rademacher random variables $\xi_1,...,\xi_n$. Then, following from (\ref{eq22}), we have
$$\xi^\top(v_I - \theta_I^*) \leq \textup{poly}(\log n) \tilde \xi^\top(v^I -\theta^{*,I}).$$
Hence, the desired inequality holds. $\hfill \square$ \\

\begin{lemma}
\label{lemma8}
Let $\delta \in \mathbb{R}^{N^d}$ with $\Delta^2(\delta) \leq t^2$. Then
$$\Vert\Pi\delta\Vert_\infty \leq \frac{t^2}{N^d} + \frac{t}{\sqrt{N^d}}.$$
\end{lemma}
\noindent \textit{Proof.} Notice that $\text{Ker}(D) = \text{span}(1_{N^d})$, then $\Pi\delta = \delta^\top v \cdot v,$ where $v= \frac{1}{\sqrt{N^d}}1_{N^d}$. Hence
\begin{equation}
\label{eq26}
    \Vert\Pi\delta\Vert_\infty = \Vert\delta^\top v \cdot v\Vert_\infty \leq |\delta^\top v| \cdot \Vert v \Vert_\infty = \frac{|\delta^\top v|}{\sqrt{N^d}}.
\end{equation}
Now, 
\begin{equation}
\label{eq27}
\begin{split}
    |\delta^\top v| &\leq \sum_{i=1}^{N^d}|\delta_i||v_i|\\
      &= \sum_{i=1}^{N^d}|\delta_i||v_i|1_{\{|\delta_i| > L\}}\ + \sum_{i=1}^{N^d}|\delta_i||v_i|1_{\{|\delta_i| \leq L\}}\ \\
      &\leq \Vert v\Vert_\infty\sum_{i=1}^{N^d} |\delta_i|1_{\{|\delta_i| > L\}} + 
      \Vert v\Vert \left(\sum_{i=1}^{N^d} \delta_i^2 1_{\{|\delta_i| \leq L\}}\right)^{1/2}\\
      &\leq \frac{t^2}{\sqrt{N^d}} + t,
\end{split}
\end{equation}
where the first inequality follows from the triangle inequality, the second from Hölder's and Cauchy Schwarz inequalities. The claim follows combining (\ref{eq26}) with (\ref{eq27}). $\hfill \square$\\

\begin{lemma}
\label{lemma9} Under Assumptions \ref{as1}-\ref{as5}, for $N \asymp n^{1/d}$, we have that 
$$SGW\left(\left\{\delta:\delta \in \tilde S , \Delta^2(\delta)\leq c_1t^2\right\}\right) \leq C_d\left(\frac{c_1t^2}{\sqrt{n}} +\sqrt{c_1}t\right) + Vn^{1-1/d}\textup{poly}(\log n).$$
\end{lemma}
\noindent \textit{Proof.} Recall that the projection on $\text{Ker}(D)^\perp$, $D^\dagger D = I - \Pi$, which yields
$$\tilde\xi^\top \delta = \tilde\xi^\top \Pi \delta + \tilde\xi^\top D^\dagger D\delta.$$
Then we have
\begin{equation}
\label{eq28}
\begin{split}
\mathbb{E}\left[\sup_{\delta \in \tilde S: \Delta^2(\delta)\leq t^2}\tilde\xi^\top\delta\right] &\leq \mathbb{E}\left[\sup_{\delta \in \tilde S: \Delta^2(\delta)\leq t^2}\tilde\xi^\top \Pi \delta\right] + \mathbb{E}\left[\sup_{\delta \in \tilde S: \Delta^2(\delta)\leq t^2}\tilde\xi^\top D^\dagger D\delta\right] \\
&=: \ T_1 + T_2.
\end{split}
\end{equation}
We first bound $T_1$. Notice that $\Pi$ is idempotent, i.e., $\Pi^2 = \Pi$, thus
\begin{equation}
\label{eq29}
\begin{split}
     \tilde\xi^\top \Pi \delta &= \tilde\xi^\top \Pi\Pi \delta  \\
     &\leq \Vert\tilde\xi^\top \Pi\Vert_1\Vert\Pi \delta \Vert_\infty \\ & = \bigg\vert \sum_{j=1}^{N^d} \tilde\xi_j \bigg\vert \cdot \Vert\Pi \delta \Vert_\infty \\
      &= \bigg\vert \sum_{j=1}^{N^d} \frac{\tilde\xi_j}{ N^{d/2} }  \bigg\vert \cdot  N^{d/2}\Vert\Pi \delta \Vert_\infty \\
\end{split}
\end{equation}
where the inequality follow from Hölder's inequality, and $v= \frac{1}{\sqrt{N^d}}1_{N^d}$. Then, 
\begin{equation}
\label{eq30}
     T_1  \leq \mathbb{E}\left[\bigg\vert \sum_{j=1}^{N^d} \frac{\tilde\xi_j}{ N^{d/2} }  \bigg\vert\right]\left(\frac{c_1t^2}{\sqrt{N^d}} +\sqrt{c_1}t\right) \leq C_d\left(\frac{c_1t^2}{\sqrt{n}} +\sqrt{c_1}t\right)
\end{equation}
for some positive constant $C_d$.\\ \\
Next, we bound $T_2$. Notice that 
\begin{equation}
\label{eq31}
\begin{split}
T_2 &\leq \mathbb{E}\left[\sup_{\delta \in \tilde S: \Delta^2(\delta)\leq t^2} \Vert\tilde\xi^\top D^\dagger\Vert_\infty \Vert D\delta\Vert_1 \right] \\
&\leq Vn^{1-1/d}\textup{poly}(\log n)\mathbb{E}\left[\Vert\tilde\xi^\top D^\dagger\Vert_\infty \right],
\end{split}
\end{equation}
thus we only need to bound $\mathbb{E}\left[\Vert\tilde\xi^\top D^\dagger\Vert_\infty \right]$. From Section 3 in \citet{Huetter16}, we write $D^\dagger = [s_1,...,s_m]$, and $\max_{j = 1,...,m}\Vert s_j\Vert_2$ is bounded above by bounded by $(\log n)^{1/2}$ for $d = 2$ or by a constant for $d>2$. Then,
\begin{equation}
\label{eq32}
\begin{split}
    \mathbb{E}\left[\Vert\tilde\xi^\top D^\dagger\Vert_\infty\right] &= \mathbb{E}\left[\max_{j = 1,...,m} |s_j^\top \tilde\xi|\right] \\
    &= \mathbb{E}\left[\max_{j = 1,...,m} |\tilde s_j^\top \tilde\xi|\right]\max_{j = 1,...,m} \Vert s_j\Vert_2, 
\end{split}
\end{equation}
where $$\tilde s_j = \frac{s_j}{\max_{j = 1,...,m} \Vert s_j\Vert_2}.$$
Moreover, $\tilde s_j^\top \tilde \xi$ is also sub-Gaussian but with parameter at most 1. Combining (\ref{eq31}) with (\ref{eq32}), we obtain that
$$T_2 \leq C_d V n^{1-1/d}\textup{poly}(\log n),$$
for some positive constant $C_d$. The conclusion follows then. $\hfill\square$ \\

\begin{theorem}
\label{theorem10}
Suppose that 
$$2SGW\left(\{\delta\in \tilde S :\Delta^2(\delta^I) \leq c_1\eta^2\}\right)\leq \kappa(\eta),$$
for a function $\kappa:\mathbb{R}\rightarrow\mathbb{R}$. Then for all $\eta > 0$ we have that
$$\mathbb{P}\left(\Delta^2(\hat \delta)>\eta^2\right) \leq \frac{\textup{poly}(\log n)\kappa(\eta)}{c_\tau \eta^2}  + \frac{V n^{1-1/d}\textup{poly}(\log n)}{c_\tau \eta^2}  ,$$
where $c_\tau$ is the constant from Lemma \ref{lemma2}. Furthermore, if $\{r_n\}$ is a sequence such that 
$$\lim_{t\rightarrow \infty} \sup_n \left[\frac{\textup{poly}(\log n)\kappa(t r_n n^{1/2})}{t^2r_n^2n} + \frac{V n^{1-1/d}\textup{poly}(\log n)}{t^2r_n^2n }\right]  \rightarrow 0,$$
then 
$$\frac{1}{n}\Delta^2(\hat \theta -\theta^*) = O_{\mathbb{P}}(r_n^2).$$
\end{theorem}
\noindent \textit{Proof.} Let $\hat \delta = \hat \theta -\theta^*$ and suppose that
\begin{equation}
\label{eq33}
    \frac{1}{n}\delta^2 > \frac{\eta^2}{n}.
\end{equation}
Next, let $q^2 = \Delta^2(\hat \delta)$. Then define $g:[0,1]\rightarrow \mathbb{R}$ as $g(t) = \Delta^2(t\hat\delta)$. Clearly, $g$ is a continuous function with $g(0) = 0$ and $g(1) = q^2$. Therefore, there exists $t_{\hat\delta}$ such that $g(t_{\hat\delta})=\eta^2$.
Hence, letting $\tilde \delta = t_{\hat\delta}$ we observe that by the basic inequality $\hat M(\theta^*+\tilde \delta) \leq 0, \delta \in S$ by convexity of $S$, and $\Delta^2(\tilde \delta)=\eta^2$ by construction. This implies, along with Lemma \ref{lemma2}, that
$$\sup_{v\in S:\Delta^2(v-\theta^*)\leq \eta^2}M(v) - \hat M(v) \geq M(\theta^*+\tilde\delta) - \hat M(\theta^*+\tilde\delta)\geq M(\theta^*+\tilde \delta) \geq c_\tau \eta^2.$$
Therefore, combing the results of Lemma \ref{lemma4}-\ref{lemma7}, we have
\begin{align*}
    \mathbb{P}\left\{\Delta^2(\hat \delta) > \eta^2\right\} &\leq \mathbb{P}\left\{\sup_{v\in S:\Delta^2(v-\theta^*)\leq \eta^2}M(v) - \hat M(v) \geq  c_\tau \eta^2 \right\}\\
    &\leq \frac{1}{c_\tau \eta^2}\mathbb{E}\left\{\sup_{v\in S:\Delta^2(v-\theta^*)\leq \eta^2}M(v) - \hat M(v)\right\} \\
    &\leq \frac{1}{c_\tau \eta^2}\mathbb{E}\left\{\sup_{v\in \tilde S:\Delta^2(v-\theta^*)\leq c_1\eta^2}M(v) - \hat M(v)\right\} \\
    &\leq \frac{2\textup{poly}(\log n)}{c_\tau \eta^2}SGW\left(\{\delta\in \tilde S :\Delta^2(\delta^I) \leq c_1\eta^2\}\right)+ \\
     &\,\,\,\,\,\,\, \,\,   \frac{V n^{1-1/d}\textup{poly}(\log n)}{c_\tau \eta^2}\\
    &\leq \frac{\textup{poly}(\log n)\kappa(\eta)}{c_\tau \eta^2} + \frac{V n^{1-1/d}\textup{poly}(\log n)}{c_\tau \eta^2},
\end{align*}
where the second inequality follows from Markov's inequality. This completes the proof. $\hfill \square$

\subsection*{F.3 \ \ Proof of Theorem \ref{theorem1}}
\noindent \textit{Proof.}
The claim follows immediately from Lemmas \ref{lemma7} and \ref{lemma9} and Theorem \ref{theorem10} by setting
$$r_n \asymp n^{-1/2d}\textup{poly}(\log n).$$ $\hfill \square$

\section*{Appendix G. Theorem \ref{theorem2}}
\subsection*{G.1 \ \ Auxiliary lemmas for Proof of Theorem \ref{theorem2}}
Throughout we assume that Assumptions \ref{as1}-\ref{as5} hold, and all notations follow the same as in the proof of Theorem 1.

\begin{lemma}
\label{lemma11} 
Let $\epsilon \in (0,1)$, then there exists a choice
$$\lambda = \begin{cases}\Theta\left\{\log n\right\} & \text{ for } d = 2,\\
\Theta\left\{(\log n)^{1/2}\right\} & \text{ for } d > 2,\\
\end{cases}$$
such that for a constant $C_0 > 0$, we have that, with probability at least $1-\epsilon/4$,
$$\kappa(\hat \theta - \theta^*) \in \mathcal{A},$$
with 
$$\mathcal{A}:= \left\{\delta : \Vert \nabla_G \delta\Vert_1 \leq C_0\left(\Vert \nabla_G \theta^*\Vert_1 + \frac{R_1}{R_2}\left[\frac{c_1\Delta^2(\delta)}{n^{1/2}}+ \sqrt{c_1}\Delta(\delta)\right]\right)\right\}$$
for all $\kappa \in [0,1],$
where
$$R_1 = C_d \log\left\{\frac{n}{\epsilon}\right\}^{1/2},$$
$$R_2 =\begin{cases} C_d(\log n)^{1/2}\left[\log\left\{\frac{c_k n}{\epsilon}\right\}\right]^{1/2} & \text{ for } d =2,\\
C_d\left[\log\left\{\frac{c_k n}{\epsilon}\right\}\right]^{1/2}& \text{ for } d > 2,
\end{cases}$$
and $C_0, C_d$ are positive constants. 
\end{lemma}
\noindent \textit{Proof.} Pick $\kappa \in [0,1]$ fixed, and let $\tilde \delta = \kappa(\hat \theta - \theta^*)$. Then by the optimality of $\hat \theta$ and the convexity of (\ref{eq8}), we have that
$$\sum_{i=1}^n \rho_\tau (y_i - \tilde \theta_i) + \lambda \Vert \nabla_G \tilde \theta\Vert_1 \leq \sum_{i=1}^n \rho_\tau (y_i - \theta^*_i) + \lambda \Vert \nabla_G \theta^*\Vert_1,$$
where $\tilde \theta =\theta^* + \tilde \delta$. Then as in the proof of Lemma 3 from \citet{Belloni11},
\begin{equation}
\label{eq34}
    0 \leq \lambda\left[\Vert \nabla_G\theta^*\Vert_1 - \Vert\nabla_G \tilde \theta\Vert_1\right] + (\tilde \theta -\theta^*)^\top a^*,
\end{equation}
where $a_i^* = \tau - 1\{y_i \leq \theta_i^*\}$ for $i = 1,...,n$.
\\ \\
Next, we bound the second term of the right hand of (\ref{eq34}). From Lemma \ref{lemma7}, we know it is sufficient to bound 
$$\tilde{a}^\top(\tilde\theta^I-\theta^{*,I}),$$
where 
$$\tilde{a}_j = \left\{\max_{u\in I}|C(u)|\right\}^{-1/2}\sum_{l \in I_j} a^*_l.$$
Now, 
\begin{equation}
\label{eq35}
\begin{split}
    \tilde{a}^\top(\tilde\theta^I-\theta^{*,I}) & = \tilde{a}^\top \Pi(\tilde\theta^I-\theta^{*,I}) + \tilde{a}^\top D^\dagger D(\tilde\theta^I-\theta^{*,I}) \\
    &=: A_1 + A_2
\end{split}
\end{equation}
From Lemmas \ref{lemma6}, \ref{lemma8} and (\ref{eq29}), we obtain
\begin{equation}
\label{eq36}
\begin{split}
    A_1 &\leq \Vert \tilde a\Vert_\infty\left(\frac{\Delta^2(\tilde\theta^I-\theta^{*,I})}{n^{1/2}}+ \Delta(\tilde\theta^I-\theta^{*,I})\right) \\ 
    &\leq \Vert \tilde a\Vert_\infty\left(\frac{c_1\Delta^2(\tilde\theta-\theta^{*})}{n^{1/2}}+ \sqrt{c_1}\Delta(\tilde\theta-\theta^{*})\right)
\end{split}
\end{equation}
To bound $A_2$, we use the result of Lemma \ref{lemma6} to obtain
\begin{equation}
\label{eq37}
\begin{split}
A_2 &\leq \Vert (D^\dagger)^\top \tilde a\Vert_\infty\left[\Vert D \tilde \theta^I\Vert_1 + \Vert D\theta^{*,I}\Vert_1\right] \\
&\leq \Vert (D^\dagger)^\top \tilde a\Vert_\infty\left[\Vert \nabla_G \tilde \theta\Vert_1 + \Vert \nabla_G \theta^*\Vert_1\right]
\end{split}
\end{equation}
Since $a^*$ is Bernoulli with parameter $\tau$ and is thus sub-Gaussian with parameter $\frac{1}{4}$, $\tilde a$ is also sub-Gaussian. As in the proof of Theorem 2 from \citet{Huetter16}, we have that the following two inequalities hold simultaneously on an event of probability at least $1 - 2\epsilon$,
$$\Vert \tilde a\Vert_\infty \leq R_1 : =C_d \log\left\{\frac{n}{\epsilon}\right\}^{1/2}, \ \ \ \ \Vert (D^\dagger)^\top \tilde a\Vert_\infty \leq R_2 := \begin{cases} C_d(\log n)^{1/2}\left[\log\left\{\frac{c_k n}{\epsilon}\right\}\right]^{1/2} & \text{ for } d =2,\\
C_d\left[\log\left\{\frac{c_k n}{\epsilon}\right\}\right]^{1/2}& \text{ for } d >2,
\end{cases}$$
for some constant $c_k$. \\ \\
Then, with probability at least $1 - 2\epsilon$,
By choosing $\lambda = 2R_2$, we obtain
$$\kappa(\hat \theta -\theta^*) \in \mathcal{A} := \left\{\delta : \Vert \nabla_G \delta\Vert_1 \leq C_0\left(\Vert \nabla_G \theta^*\Vert_1 + \frac{R_1}{R_2}\left[\frac{c_1\Delta^2(\delta)}{n^{1/2}}+ \sqrt{c_1}\Delta(\delta)\right]\right)\right\},$$
for some positive constant $C_0$. $\hfill \square$

\subsection*{G.2 \ \ Proof of Theorem \ref{theorem2}}
\noindent \textit{Proof.} Let $\epsilon \in (0,1)$. By Lemma \ref{lemma11} we can suppose that the following event
\begin{equation}
\label{eq38}
    \Omega  = \left\{\kappa(\hat \theta -\theta^*) \in \mathcal{A}, \ \forall \kappa \in [0,1]\right\} 
\end{equation}
happen with probability at least $1-\epsilon/2$ with $\mathcal{A}$ as in Lemma \ref{lemma11}. Then,
$$\mathbb{P}\left\{\Delta^2(\hat\delta) > \eta^2 \right\} \leq \mathbb{P}\left[\left\{\Delta^2(\hat\delta) > \eta^2 \right\}\cap \Omega\right] + \frac{\epsilon}{2}.$$
Now suppose that the event 
$$\left\{\Delta^2(\hat\delta) > \eta^2 \right\}\cap \Omega$$
holds. As in the proof of Theorem \ref{theorem10}, there exists $\tilde \delta = t_{\hat\delta}\hat\delta$ with $t_{\hat\delta} \in [0,1]$ such that $\tilde \delta \in \mathcal{A}$, $\Delta^2(\tilde \delta) = \eta^2$. Hence, by the basic inequality, 
$$\hat M (\theta^* + \tilde\delta) + \lambda \left[\Vert \nabla_G(\theta^* + \tilde\delta)\Vert_1 - \Vert\nabla_G\theta^*\Vert_1\right] \leq 0.$$
Then, 
\begin{align*}
    \sup_{\delta \in \mathcal{A}, \Delta^2(\delta) \leq \eta^2} \left[M(\theta^*+\delta) - \hat M(\theta^*+\delta) + \lambda \left\{\Vert\nabla_G\theta^*\Vert_1 -\Vert\nabla_G(\theta^* + \tilde\delta)\Vert_1 \right\}\right]&\geq M(\theta^* +  \delta) \\
    &\geq c_\tau \eta^2,
\end{align*}
where the second inequality follows from Lemma \ref{lemma2}. Therefore, 
\begin{equation}
\label{eq39}
\begin{split}
\mathbb{P}\left[\left\{\Delta^2(\hat\delta) > \eta^2 \right\}\cap \Omega\right] &\leq \mathbb{P}\Bigg(\Bigg\{\sup_{\delta \in \mathcal{A}, \Delta^2(\delta) \leq \eta^2} \Big[M(\theta^*+\delta) - \hat M(\theta^*+\delta)  \\
& \ \ \ \ \ \ \ \ \ \ \ \ \ \ \ \ \ \ \ \ \ \ \ \ \ \ \  + \lambda \Big\{\Vert\nabla_G\theta^*\Vert_1 -\Vert\nabla_G(\theta^* + \delta)\Vert_1 \Big\}\Big] \geq c_\tau\eta^2\Bigg\} \cap \Omega\Bigg) \\
&\leq \frac{1}{c_\tau \eta^2}\mathbb{E}\Bigg(\textbf{1}_\Omega\sup_{\delta \in \mathcal{A}, \Delta^2(\delta) \leq \eta^2} \Big[M(\theta^*+\delta) - \hat M(\theta^*+\delta) \\
& \ \ \ \ \ \ \ \ \ \ \ \ \ \ \ \ \ \ \ \ \ \ \ \ \ \ \ \ \ \ \ \ \ \ \ \ + \lambda \Big\{\Vert\nabla_G\theta^*\Vert_1 -\Vert\nabla_G(\theta^* + \delta)\Vert_1 \Big\}\Big]\Bigg)\\
&\leq \frac{1}{c_\tau \eta^2}\mathbb{E}\left(\textbf{1}_\Omega\sup_{\delta \in \mathcal{A}, \Delta^2(\delta) \leq \eta^2} \left[M(\theta^*+\delta) - \hat M(\theta^*+\delta) \right]\right) \\
& \ \ \ \ + \frac{\lambda}{c_\tau \eta^2}\mathbb{E}\left(\textbf{1}_\Omega\sup_{\delta \in \mathcal{A}, \Delta^2(\delta) \leq \eta^2} \left[\Vert\nabla_G\theta^*\Vert_1 -\Vert\nabla_G(\theta^* + \delta)\Vert_1 \right]\right) \\
&\leq \frac{1}{c_\tau \eta^2}\mathbb{E}\left(\textbf{1}_\Omega\sup_{\delta \in \mathcal{A}, \Delta^2(\delta) \leq \eta^2} \left[M(\theta^*+\delta) - \hat M(\theta^*+\delta) \right]\right) \\
& \ \ \ \ + \frac{\lambda}{c_\tau \eta^2}\mathbb{E}\left(\textbf{1}_\Omega\sup_{\delta \in \mathcal{A}, \Delta^2(\delta) \leq \eta^2} \Vert\nabla_G\delta\Vert_1\right), \\
\end{split}
\end{equation}
where the second inequality follows from Markov’s inequality, and the last from the triangle inequality.
\\
\\
Next, define
$$\mathcal{H}(\eta) = \{\delta \in \mathcal{A}:\Delta(\delta)\leq \eta\}.$$
Hence, if $\delta \in \mathcal{H}(\eta)$ and $\Omega$ holds, then
\begin{equation}
\begin{split}
    \Vert\nabla_G\delta\Vert_1 &\leq C_0\left(\Vert \nabla_G\theta^*\Vert_1 + \frac{R_1}{R_2}\left[\frac{c_1\Delta^2(\delta)}{n^{1/2}}+ \sqrt{c_1}\Delta(\delta)\right]\right)\\
\end{split}
\end{equation}
where the inequality follow from the definition of $\mathcal{H}(\eta)$ and Lemma \ref{lemma11}.\\ \\
We now define
$$\mathcal{L}(\eta) = \left\{\delta:\Vert\nabla_G\delta\Vert_1 \leq C_0\left\{\Vert \nabla_G\theta^*\Vert_1 + \frac{R_1}{R_2}\left[\frac{c_1\Delta^2(\delta)}{n^{1/2}}+ \sqrt{c_1}\Delta(\delta)\right]\right\}, \Delta(\delta)\leq \eta\right\},$$
$$\tilde{\mathcal{L}}(\eta) = \left\{\delta:\Vert D\delta^I\Vert_1 \leq C_0\left\{\Vert \nabla_G\theta^*\Vert_1 + \frac{R_1}{R_2}\left[\frac{c_1\Delta^2(\delta)}{n^{1/2}}+ \sqrt{c_1}\Delta(\delta)\right]\right\}, \Delta(\delta^I)\leq \sqrt{c_1}\eta\right\}.$$
Then
\begin{equation}
\begin{split}
\mathbb{P}\left[\left\{\Delta^2(\hat\delta) > \eta^2 \right\}\cap \Omega\right] 
&\leq \frac{1}{c_\tau \eta^2}\mathbb{E}\left(\sup_{\delta \in  {\mathcal{L}}(\eta)} \left[M(\theta^*+\delta) - \hat M(\theta^*+\delta) \right]\right) + \frac{\lambda}{c_\tau \eta^2}\sup_{\delta \in \mathcal{L}(\eta)}\Vert\nabla_G\delta\Vert_1 \\
&\leq \frac{2\textup{poly}(\log n)}{c_\tau \eta^2}\mathbb{E}\left(\sup_{\delta \in \tilde{\mathcal{L}}(\eta)} \sum_{i=1}^{N^d} \tilde\xi_i\delta^I_i\right) + \frac{2C_0 V^* n^{1-1/d}\textup{poly}(\log n)}{c_\tau \eta^2}+\\
 &   \,\,\,\,\,\,\, \frac{2C_0\textup{poly}(\log n)}{c_{\tau}\eta^2}\frac{R_1}{R_2}\left[\frac{c_1\eta^2}{n^{1/2}}+ \sqrt{c_1}\eta\right]  +   \frac{\lambda}{c_\tau \eta^2}\sup_{\delta \in \mathcal{L}(\eta)}\Vert\nabla_G\delta\Vert_1.
\end{split}
\end{equation}
where the second inequality follows with the same argument from Lemmas \ref{lemma4}, \ref{lemma5}, and \ref{lemma7}. \\ \\
By Lemma \ref{lemma9}, we have
\begin{equation*}
\begin{split}
\mathbb{P}\left[\left\{\Delta^2(\hat\delta) > \eta^2 \right\}\cap \Omega\right] 
&\leq \frac{2\textup{poly}(\log n)}{c_\tau \eta^2}\Bigg\{C_d\left(\frac{c_1\eta^2}{n^{1/2}} +\sqrt{c_1}\eta\right)(\log n)^{1/2} \\
& \ \ \ \ +C_d C_0 \left[V^*n^{1-1/d}\textup{poly}(\log n) + \frac{R_1}{R_2}\left(\frac{c_1\eta^2}{n^{1/2}}+\sqrt{c_1}\eta\right)\right]\\
& \ \ \ \ + \frac{\lambda}{c_\tau \eta^2}C_0\left[V^*n^{1-1/d}\textup{poly}(\log n) + \frac{R_1}{R_2}\left(\frac{c_1\eta^2}{n^{1/2}}+ \sqrt{c_1}\eta\right)\right]\Bigg\}.
\end{split}
\end{equation*}
Hence given our choice of $\lambda$, by choosing $$\eta = \
c_\gamma n^{\frac{1}{2}(1-1/d)}\textup{poly}(\log n)$$
for some  $c_\gamma > 1$,
we conclude that
$$\mathbb{P}\left[\left\{\Delta^2(\hat\delta) > \eta^2 \right\}\cap \Omega\right] \leq \epsilon,$$
provided that $c_\gamma$ is large enough. $\hfill \square$


\newpage
\vskip 0.2in
\bibliography{reference}
\end{document}